\documentclass[twocolumn]{revtex4}

\usepackage{graphicx}
\usepackage{dcolumn}

\usepackage{amstext,amsmath,amssymb,amsfonts,braket}
\usepackage{bm}

\usepackage{hyperref}

\usepackage{color}
\definecolor{lightblue}{rgb}{0.2,0.2,0.7}
\definecolor{darkblue}{rgb}{0,0.25,0.5}
\definecolor{redbrown}{rgb}{0.875,0.25,0.125}
\definecolor{darkgreen}{rgb}{0,0.5,0}

\renewcommand{\b}[1]{\ensuremath{\mathbf{#1}}}
\renewcommand{\H}{\ensuremath{\text{H}}}
\renewcommand{\l}{\ensuremath{\lambda}}

\newcommand{\lr}{\ensuremath{\text{lr}}}
\newcommand{\sr}{\ensuremath{\text{sr}}}
\newcommand{\ee}{\ensuremath{\text{ee}}}

\newcommand{\HF}{\ensuremath{\text{HF}}}
\newcommand{\MP}{\ensuremath{\text{MP2}}}

\renewcommand{\d}{\ensuremath{\text{d}}}
\newcommand{\s}{\ensuremath{\text{s}}}
\newcommand{\x}{\ensuremath{\text{x}}}
\newcommand{\xc}{\ensuremath{\text{xc}}}
\renewcommand{\c}{\ensuremath{\text{c}}}
\newcommand{\Hxc}{\ensuremath{\text{Hxc}}}
\newcommand{\Hx}{\ensuremath{\text{Hx}}}

\DeclareMathOperator{\erf}{erf}
\DeclareMathOperator{\erfc}{erfc}

\newcommand{\isEquivTo}[1]{\underset{#1}{\sim}}

\makeatletter
\renewcommand\paragraph{\@startsection{paragraph}{4}{\z@}%
  {-3.25ex\@plus -1ex \@minus -.2ex}%
  {1.5ex \@plus .2ex}%
  {\normalfont\normalsize\bfseries}}
\makeatother

\begin{document}

\title{A general range-separated double-hybrid density-functional theory}

\author{Cairedine Kalai}
\author{Julien Toulouse}\email{toulouse@lct.jussieu.fr}
\affiliation{
 Laboratoire de Chimie Th\'eorique (LCT), Sorbonne Universit\'e and CNRS, F-75005 Paris, France
}

\date{April 7, 2018}

\begin{abstract}
A range-separated double-hybrid (RSDH) scheme which generalizes the usual range-separated hybrids and double hybrids is developed. This scheme consistently uses a two-parameter Coulomb-attenuating-method (CAM)-like decomposition of the electron-electron interaction for both exchange and correlation in order to combine Hartree-Fock exchange and second-order M{\o}ller-Plesset (MP2) correlation with a density functional. The RSDH scheme relies on an exact theory which is presented in some detail. Several semi-local approximations are developed for the short-range exchange-correlation density functional involved in this scheme. After finding optimal values for the two parameters of the CAM-like decomposition, the RSDH scheme is shown to have a relatively small basis dependence and to provide atomization energies, reaction barrier heights, and weak intermolecular interactions globally more accurate or comparable to range-separated MP2 or standard MP2. The RSDH scheme represents a new family of double hybrids with minimal empiricism which could be useful for general chemical applications.
\end{abstract}

\maketitle

\section{Introduction}

Over the past two decades, density-functional theory (DFT) \cite{HohKoh-PR-64} within the Kohn-Sham (KS) scheme~\cite{KohSha-PR-65} has been a method of choice to study ground-state properties of electronic systems. KS DFT is formally exact, but it involves the so-called exchange-correlation energy functional whose explicit form in terms of the electron density is still unknown. Hence, families of approximations to this quantity have been developed: semi-local approximations (local-density approximation (LDA), generalized-gradient approximations (GGAs) and meta-GGAs), hybrid approximations, and approximations depending on virtual orbitals (see, e.g., Ref.~\onlinecite{SuLiTru-JCP-16} for a recent review).

This last family of approximations includes approaches combining semi-local density-functional approximations (DFAs) with Hartree--Fock (HF) exchange and second-order M{\o}ller--Plesset (MP2) correlation, either based on a range separation of the electron-electron interaction or a linear separation. In the range-separated hybrid (RSH) variant, the Coulomb electron-electron interaction $w_\ee(r_{12})=1/r_{12}$ is decomposed as~\cite{Sav-INC-96,TouColSav-PRA-04}
\begin{equation}
w_\ee(r_{12}) = w_\ee^{\lr,\mu} (r_{12}) + w_\ee^{\sr,\mu} (r_{12}),
\label{Ewald-style}
\end{equation}
where $w_\ee^{\lr,\mu} (r_{12}) = \erf(\mu r)/r_{12}$ is a long-range interaction (written with the error function erf) and $w_\ee^{\sr,\mu} (r_{12}) = \erfc(\mu r)/r_{12}$ is the complementary short-range interaction (written with the complementary error function erfc), the decomposition being controlled by the parameter $\mu$ ($0 \leq \mu < \infty$). HF exchange and MP2 correlation can then be used for the long-range part of the energy, while a semi-local exchange-correlation DFA is used for the complementary short-range part, resulting in a method that is denoted by RSH+MP2~\cite{AngGerSavTou-PRA-05}. Among the main advantages of such an approach are the explicit description of van der Waals dispersion interactions via the long-range MP2 part (see, e.g., Ref.~\onlinecite{TayAngGalZhaGygHirSonRahLilPodBulHenScuTouPevTruSza-JCP-16}) and the fast (exponential) convergence of the long-range MP2 correlation energy with respect to the size of the one-electron basis set~\cite{FraMusLupTou-JCP-15}. On the other hand, the short-range exchange-correlation DFAs used still exhibit significant errors, such as self-interaction errors~\cite{MusTou-MP-17}, limiting the accuracy for the calculations of atomization energies or non-covalent electrostatic interactions for example.

Similarly, the double-hybrid (DH) variant~\cite{Gri-JCP-06} (see Ref.~\onlinecite{GoeGri-WIRE-14} for a review) for combining MP2 and a semi-local DFA can be considered as corresponding to a linear separation of the Coulomb electron-electron interaction~\cite{ShaTouSav-JCP-11}
\begin{equation}
w_\ee(r_{12}) = \l w_\ee (r_{12}) + (1-\l) w_\ee (r_{12}),
\label{decomplinear}
\end{equation}
where $\l$ is a parameter ($ 0 \leq \l \leq 1$). If HF exchange and MP2 correlation is used for the part of the energy associated with the interaction $\l w_\ee (r_{12})$ and a semi-local exchange-correlation DFA is used for the complementary part, then a one-parameter version of the DH approximations is obtained~\cite{ShaTouSav-JCP-11}. One of the main advantages of the DH approximations is their quite efficient reduction of the self-interaction error~\cite{SuYanMorXu-JPCA-14} thanks to their large fraction of HF exchange ($\l \approx 0.5$ or more). On the other hand, they inherit (a fraction of) the slow (polynomial) basis convergence of standard MP2~\cite{KarTarLamSchMar-JPCA-08}, and they are insufficiently accurate for the description of van der Waals dispersion interactions and need the addition of semi-empirical dispersion corrections~\cite{SchGri-PCCP-07}.

In this work, we consider range-separated double-hybrid (RSDH)~\cite{KalTou-JJJ-XX-note2} approximations which combine the two above-mentioned approaches, based on the following decomposition of the Coulomb electron-electron interaction
\begin{eqnarray}
w_\ee(r_{12}) &=& \left[ w_\ee^{\lr,\mu} (r_{12}) + \l w_\ee^{\sr,\mu} (r_{12}) \right] 
\nonumber\\
&&+ (1-\l) w_\ee^{\sr,\mu} (r_{12}),
\label{decompRSDH}
\end{eqnarray}
where, again, the energy corresponding to the first part of the interaction (in square brackets) is calculated with HF exchange and MP2 correlation, and the complementary part is treated by a semi-local exchange-correlation DFA. The expected features of such an approach are the explicit description of van der Waals dispersion interactions through the long-range part, and reduced self-interaction errors in the short-range part (and thus improved calculations of properties such as atomization energies).

The decomposition of Eq.~(\ref{decompRSDH}) is in fact a special case of the decomposition used in the Coulomb-attenuating method (CAM)~\cite{YanTewHan-CPL-04}
\begin{eqnarray}
w_\ee(r_{12}) = \left[ (\alpha + \beta) w_\ee^{\lr,\mu} (r_{12}) + \alpha w_\ee^{\sr,\mu} (r_{12}) \right] 
\nonumber\\
+ \left[ (1-\alpha - \beta) w_\ee^{\lr,\mu} (r_{12}) + (1-\alpha) w_\ee^{\sr,\mu} (r_{12}) \right],
\label{CAM}
\end{eqnarray}
with the parameters $\alpha + \beta = 1$ and $\alpha = \l$. The choice $\alpha + \beta = 1$ is known to be appropriate for Rydberg and charge-transfer excitation energies~\cite{PeaHelSalKeaLutTozHan-PCCP-06} and for reaction-barrier heights~\cite{PeaCohToz-PCCP-06}. We also expect it to be appropriate for the description of long-range van der Waals dispersion interactions. It should be noted that the CAM decomposition has been introduced in Ref.~\onlinecite{YanTewHan-CPL-04} at the exchange level only, i.e. for combining HF exchange with a semi-local exchange DFA without modifying the semi-local correlation DFA. Only recently, Cornaton and Fromager~\cite{CorFro-IJQC-14} (see also Ref.~\onlinecite{CorStoJenFro-PRA-13}) pointed out the possibility of a CAM double-hybrid approximation but remarked that the inclusion of a fraction of short-range electron-electron interaction in the MP2 part would limit the basis convergence, and preferred to develop an alternative approach which uses only the perturbation expansion of a long-range interacting wave function. Despite the expected slower basis convergence of DH approximations based on the decomposition in Eq.~(\ref{decompRSDH}) [or in Eq.~(\ref{CAM})] in comparison to the RSH+MP2 method based on the decomposition in Eq.~(\ref{Ewald-style}), we still believe it worthwhile to explore this kind of DH approximations in light of the above-mentioned expected advantages. In fact, we will show that the basis convergence of the RSDH approximations is relatively fast, owing to the inclusion of a modest fraction of short-range MP2 correlation.

The decomposition in Eq.~(\ref{decompRSDH}) has been used several times at the exchange level~\cite{ChaHea-JCP-08,LanRohHer-JPCB-08,RohHer-JCP-08,RohMarHer-JCP-09,PevTru-JPCL-11,NguDayPac-JCP-11,RefShaGovAutNeaBaeKro-PRL-12,EggWeiRefShaDauBaeKumNeaZojKro-JCTC-14,LufRefPacResRamKroPus-PRB-14,RefJaiShaNeaKro-PRB-15,LiuEggRefKroNea-JCP-17,BruProKroBre-JCP-17}. A few DH approximations including either long-range exchange or long-range correlation terms have been proposed. The $\omega$B97X-2 approximation~\cite{ChaHea-JCP-09} adds a full-range MP2 correlation term to a hybrid approximation including long-range HF exchange. The B2-P3LYP approximation~\cite{BenDisLocChaHea-JPCA-08} and the lrc-XYG3 approximation~\cite{ZhaXu-JPCL-13} add a long-range MP2 correlation correction to a standard DH approximation including full-range HF exchange. Only in Ref.~\onlinecite{GarBulHenScu-PCCP-15} the decomposition in Eq.~(\ref{decompRSDH}) was consistently used at the exchange and correlation level, combining a pair coupled-cluster doubles approximation with a semi-local exchange-correlation DFA, in the goal of describing static correlation. However, the formulation of the exact theory based on the decomposition in Eq.~(\ref{decompRSDH}), as well as the performance of the MP2 and semi-local DFAs in this context, have not been explored. This is what we undertake in the present work.

The paper is organized as follows. In Section~\ref{sec:theory}, the theory underlying the RSDH approximations is presented, and approximations for the corresponding short-range correlation density functional are developed. Computational details are given in Section~\ref{sec:comp}. In Section~\ref{sec:results}, we give and discuss the results, concerning the optimization of the parameters $\mu$ and $\l$ on small sets of atomization energies (AE6 set) and reaction barrier heights (BH6 set), the study of the basis convergence, and the tests on large sets of atomization energies (AE49 set), reaction barrier heights (DBH24 set), and weak intermolecular interactions (S22 set). Section~\ref{sec:conclusion} contains conclusions and future work prospects. Finally, the Appendix contains the derivation of the uniform coordinate scaling relation and the Coulomb/high-density and short-range/low-density limits of the short-range correlation density functional involved in this work. Unless otherwise specified, Hartree atomic units are tacitly assumed throughout this work.

\section{Range-separated double-hybrid density-functional theory}
\label{sec:theory}

\subsection{Exact theory}

The derivation of the RSDH density-functional theory starts from the universal density functional~\cite{Lev-PNAS-79},
\begin{equation}
F[n] = \min_{\Psi \to n} \bra{\Psi} \hat{T} + \hat{W}_\ee \ket{\Psi},
\label{LL-functional}
\end{equation}
where $\hat{T}$ is the kinetic-energy operator, $\hat{W}_\ee$ the Coulomb electron-electron repulsion operator, and the minimization is done over normalized antisymmetric multideterminant wave functions $\Psi$ giving a fixed density $n$. The universal density functional is then decomposed as 
\begin{equation}
F[n] = F^{\mu,\lambda}[n] + \bar{E}_\Hxc^{\sr,\mu,\l}[n],
\label{LL-decomposition}
\end{equation}
where $F^{\mu,\lambda}[n]$ is defined as
\begin{equation}
F^{\mu,\lambda}[n] = \min_{\Psi\to n} \bra{\Psi} \hat{T} + \hat{W}_\ee^{\lr,\mu} + \l \hat{W}_\ee^{\sr,\mu} \ket{\Psi}.
\label{LL-decomposition2}
\end{equation}
In Eq.~(\ref{LL-decomposition2}), $\hat{W}_\ee^{\lr,\mu}$ is the long-range electron-electron repulsion operator and $\l \hat{W}_\ee^{\sr,\mu}$ is the short-range electron-electron repulsion operator scaled by the constant $\l$, with expressions:
\begin{align}
\hat{W}_\ee^{\lr,\mu} = \frac{1}{2} \iint w_\ee^{\lr,\mu} (r_{12}) \hat{n}_{2}(\textbf{r}_{1},\textbf{r}_{2}) \d\b{r}_{1} \d\b{r}_{2},
\end{align}
\begin{align}
\hat{W}_\ee^{\sr,\mu} = \frac{1}{2} \iint w_\ee^{\sr,\mu} (r_{12}) \hat{n}_{2}(\textbf{r}_{1},\textbf{r}_{2}) \d\b{r}_{1} \d\b{r}_{2},
\end{align}
where $\hat{n}_{2}(\textbf{r}_{1},\textbf{r}_{2})=\hat{n}(\b{r}_1) \hat{n}(\b{r}_2) - \delta(\b{r}_1-\b{r}_2) \hat{n}(\b{r}_1)$ is the pair-density operator, written with the density operator $\hat{n}(\b{r})$. Equation~(\ref{LL-decomposition}) defines the complement short-range Hartree-exchange-correlation density functional $\bar{E}_\Hxc^{\sr,\mu,\l}[n]$ depending on the two parameters $\mu$ and $\l$. It can itself be decomposed as
\begin{equation}
\bar{E}_\Hxc^{\sr,\mu,\l}[n] = E_\H^{\sr,\mu,\l}[n] + \bar{E}_{\xc}^{\sr,\mu,\l}[n],
\end{equation}
where $E_\H^{\sr,\mu,\l}[n]$ is the short-range Hartree contribution,
\begin{eqnarray}
E_\H^{\sr,\mu,\l}[n] = (1-\l) \times \;\;\;\;\;\;\;\;\;\;\;\;\;\;\;\;\;\;\;\;\;\;\;
\nonumber\\
\frac{1}{2} \iint  w_\ee^{\sr,\mu} (r_{12}) n(\b{r}_{1}) n(\b{r}_{2}) \d\b{r}_{1} \d\b{r}_{2},
\end{eqnarray}
and $\bar{E}_{\xc}^{\sr,\mu,\l}[n]$ is the short-range exchange-correlation contribution. 

The exact ground-state electronic energy of a $N$-electron system in the external nuclei-electron potential $v_{\text{ne}}(\b{r})$ can be expressed as
\begin{eqnarray}
E &=& \min_{n\to N} \Bigl\{ F[n] + \int v_{\text{ne}}(\b{r}) n(\b{r}) \d\b{r} \Bigl\}
\nonumber\\
 &=& \min_{n\to N} \Bigl\{ F^{\mu,\l}[n] + \bar{E}_\Hxc^{\sr,\mu,\l}[n] + \int v_{\text{ne}}(\b{r}) n(\b{r}) \d\b{r} \Bigl\}
\nonumber\\
 &=& \min_{\Psi\to N} \Bigl\{ \bra{\Psi} \hat{T} + \hat{W}_\ee^{\lr,\mu} + \l \hat{W}_\ee^{\sr,\mu} + \hat{V}_\text{ne} \ket{\Psi}
\nonumber\\
&& \;\;\;\;\;\;\;\;\;\; + \bar{E}_\Hxc^{\sr,\mu,\l}[n_{\Psi}] \Bigl\},
\label{GS-energy1}
\end{eqnarray}
where $n\to N$ refers to $N$-representable densities, $\Psi\to N$ refers to $N$-electron normalized antisymmetric multideterminant wave functions, and $n_{\Psi}$ denotes the density coming from $\Psi$, i.e. $n_{\Psi}(\b{r}) = \bra{\Psi} \hat{n}(\b{r}) \ket{\Psi}$. In Eq.~(\ref{GS-energy1}), the last line was obtained by using the expression of $F^{\mu,\l}[n]$ in Eq.~(\ref{LL-decomposition2}), introducing the nuclei-electron potential operator $\hat{V}_\text{ne} = \int v_{\text{ne}}(\b{r}) \hat{n}(\b{r}) \d\b{r}$, and recomposing the two-step minimization into a single one, i.e. $\min_{n\to N} \min_{\Psi\to n} = \min_{\Psi\to N}$. The minimizing wave function $\Psi^{\mu,\l}$ in Eq.~(\ref{GS-energy1}) satisfies the corresponding Euler-Lagrange equation, leading to the Schr\"odinger-like equation
\begin{eqnarray}
\Bigl( \hat{T} + \hat{W}_\ee^{\lr,\mu} + \l \hat{W}_\ee^{\sr,\mu} + \hat{V}_\text{ne} \;\;\;\;\;\;\;\;\;\;\;\;\;\;\;\;\;\;
\nonumber\\
\;\;\;\;\;+ \hat{V}_\Hxc^{\sr,\mu,\l}[n_{\Psi^{\mu,\l}}] \Bigl) \ket{\Psi^{\mu,\l}}
= {\cal E}^{\mu,\l} \ket{\Psi^{\mu,\l}}, \;
\label{EL1}
\end{eqnarray}
where ${\cal E}^{\mu,\l}$ is the Lagrange multiplier associated with the normalization constraint of the wave function. In Eq.~(\ref{EL1}), $\hat{V}_\Hxc^{\sr,\mu,\l}[n] = \int v_\Hxc^{\sr,\mu,\l}(\b{r}) \hat{n}(\b{r}) \d \b{r}$ is the complement short-range Hartree-exchange-correlation potential operator with $v_\Hxc^{\sr,\mu,\l}(\b{r}) = \delta \bar{E}_\Hxc^{\sr,\mu,\l}[n]/\delta n(\b{r})$. Equation~(\ref{EL1}) defines an effective Hamiltonian $\hat{H}^{\mu,\l} = \hat{T} + \hat{V}_\text{ne} + \hat{W}_\ee^{\lr,\mu} + \l \hat{W}_\ee^{\sr,\mu} + \hat{V}_\Hxc^{\sr,\mu,\l} [n_{\Psi^{\mu,\l}}]$ that must be solved iteratively for its ground-state multideterminant wave function $\Psi^{\mu,\l}$ which gives the exact ground-state density and the exact ground-state energy via Eq.~(\ref{GS-energy1}), independently of $\mu$ and $\l$.

We have therefore defined an exact formalism combining a wave-function calculation with a density functional. This formalism encompasses several important special cases:\\[-0.2cm]

$\bullet$ $\mu = 0$ and $\l = 0$. In Eq.~(\ref{GS-energy1}), the electron-electron operator vanishes, $\hat{W}_\ee^{\lr,\mu=0} + 0 \times \hat{W}_\ee^{\sr,\mu=0} = 0$, and the density functional reduces to the KS Hartree-exchange-correlation density functional, $\bar{E}_\Hxc^{\sr,\mu=0,\l=0}[n]=E_\Hxc[n]$, so that we recover standard KS DFT
\begin{eqnarray}
E &=& \min_{\Phi\to N} \Bigl\{ \bra{\Phi} \hat{T} + \hat{V}_\text{ne} \ket{\Phi}  + E_\Hxc[n_{\Phi}] \Bigl\},
\label{KS}
\end{eqnarray}
where $\Phi$ is a single-determinant wave function.\\

$\bullet$ $\mu \to \infty$ or $\l = 1$. In Eq.~(\ref{GS-energy1}), the electron-electron operator reduces to the Coulomb interaction $\hat{W}_\ee^{\lr,\mu\to\infty} + \l \hat{W}_\ee^{\sr,\mu\to \infty} = \hat{W}_\ee$ or $\hat{W}_\ee^{\lr,\mu} + 1 \times \hat{W}_\ee^{\sr,\mu} = \hat{W}_\ee$, and the density functional vanishes, $\bar{E}_\Hxc^{\sr,\mu\to \infty,\l}[n]= 0$ or $\bar{E}_\Hxc^{\sr,\mu,\l=1}[n]= 0$, so that we recover standard wave-function theory
\begin{eqnarray}
E &=& \min_{\Psi\to N} \bra{\Psi} \hat{T} + \hat{W}_\ee + \hat{V}_\text{ne}  \ket{\Psi}.
\end{eqnarray}

$\bullet$ $0 < \mu <\infty$ and $\l = 0$. In Eq.~(\ref{GS-energy1}), the electron-electron operator reduces to the long-range interaction $\hat{W}_\ee^{\lr,\mu} + 0 \times \hat{W}_\ee^{\sr,\mu} = \hat{W}_\ee^{\lr,\mu}$, and the density functional reduces to the usual short-range density functional, $\bar{E}_\Hxc^{\sr,\mu,\l=0}[n]=\bar{E}_\Hxc^{\sr,\mu}[n]$, so that we recover range-separated DFT~\cite{Sav-INC-96,TouColSav-PRA-04}
\begin{eqnarray}
E &=& \min_{\Psi\to N} \Bigl\{ \bra{\Psi} \hat{T} + \hat{W}_\ee^{\lr,\mu} + \hat{V}_\text{ne} \ket{\Psi}
+ \bar{E}_\Hxc^{\sr,\mu}[n_{\Psi}] \Bigl\}.
\nonumber\\
\end{eqnarray}

$\bullet$ $\mu=0$ and $0 < \l < 1$. In Eq.~(\ref{GS-energy1}), the electron-electron operator reduces to the scaled Coulomb interaction $\hat{W}_\ee^{\lr,\mu=0} + \l \hat{W}_\ee^{\sr,\mu=0} =  \l \hat{W}_\ee$, and the density functional reduces to the $\l$-complement density functional, $\bar{E}_\Hxc^{\sr,\mu=0,\l}[n]= \bar{E}_\Hxc^{\l}[n]$, so that we recover the multideterminant extension of KS DFT based on the linear decomposition of the electron-electron interaction~\cite{ShaTouSav-JCP-11,ShaSavJenTou-JCP-12}
\begin{eqnarray}
E &=& \min_{\Psi\to N} \Bigl\{ \bra{\Psi} \hat{T} + \l \hat{W}_\ee + \hat{V}_\text{ne} \ket{\Psi}
+ \bar{E}_\Hxc^{\l}[n_{\Psi}] \Bigl\}.
\nonumber\\
\end{eqnarray}

\subsection{Single-determinant approximation}
\label{sec:sd}

As a first step, we introduce a single-determinant approximation in Eq.~(\ref{GS-energy1}),
\begin{eqnarray}
E^{\mu,\l}_0 &=& \min_{\Phi \rightarrow N} \Bigl\{ \bra{\Phi} \hat{T} + \hat{W}_\ee^{\lr,\mu} + \l \hat{W}_\ee^{\sr,\mu} + \hat{V}_\text{ne} \ket{\Phi} 
\nonumber\\
&& + \bar{E}_\Hxc^{\sr,\mu,\l}[n_{\Phi}] \Bigl\},
\label{RSH-1}
\end{eqnarray}
where the search is over $N$-electron normalized single-determinant wave functions. The minimizing single determinant $\Phi^{\mu,\l}$ is given by the HF- or KS-like equation
\begin{eqnarray}
\Bigl( \hat{T} + \hat{V}_\text{ne} + \hat{V}_{\Hx,\HF}^{\lr,\mu}[\Phi^{\mu,\l}] + \l \hat{V}_{\Hx,\HF}^{\sr,\mu}[\Phi^{\mu,\l}] 
\nonumber\\
+ \hat{V}_\Hxc^{\sr,\mu,\l}[n_{\Phi^{\mu,\l}}] \Bigl) \ket{\Phi^{\mu,\l}} = {\cal E}_0^{\mu,\l} \ket{\Phi^{\mu,\l}},
\label{EL2}
\end{eqnarray}
where $\hat{V}_{\Hx,\HF}^{\lr,\mu}[\Phi^{\mu,\l}]$ and $\hat{V}_{\Hx,\HF}^{\sr,\mu}[\Phi^{\mu,\l}]$ are the long-range and short-range HF potential operators constructed with the single determinant $\Phi^{\mu,\l}$, and ${\cal E}_0^{\mu,\l}$ is the Lagrange multiplier associated with the normalization constraint. Equation~(\ref{EL2}) must be solved self-consistently for its single-determinant ground-state wave function $\Phi^{\mu,\l}$. Note that, due to the single-determinant approximation, the density $n_{\Phi^{\mu,\l}}$ is not the exact ground-state density and the energy in Eq.~(\ref{RSH-1}) is not the exact ground-state energy and depends on the parameters $\mu$ and $\l$. It can be rewritten in the form
\begin{eqnarray}
E^{\mu,\l}_0 = \bra{\Phi^{\mu,\l}} \hat{T} + \hat{V}_{\text{ne}} \ket{\Phi^{\mu,\l}} + E_\H[n_{\Phi^{\mu,\l}}] \;\;\;\;\;\;\;\;
\nonumber\\
+ E_{\x,\HF}^{\lr,\mu}[\Phi^{\mu,\l}] + \l E_{\x,\HF}^{\sr,\mu}[\Phi^{\mu,\l}] + \bar{E}_\xc^{\sr,\mu,\l}[n_{\Phi^{\mu,\l}}], \;\;
\label{Emul0}
\end{eqnarray}
where $E_\H[n] = (1/2) \iint w_\ee(r_{12}) n(\b{r}_1) n(\b{r}_2) \d \b{r}_1 \d \b{r}_2$ is the standard Hartree energy with the Coulomb electron-electron interaction, and $E_{\x,\HF}^{\lr,\mu}$ and $E_{\x,\HF}^{\sr,\mu}$ are the long-range and short-range HF exchange energies. For $\mu = 0$ and $\l = 0$, we recover standard KS DFT, while for $\mu \rightarrow \infty$ or $\l = 1$ we recover standard HF theory. For intermediate values of $\mu$ and $\l$, this scheme is very similar to the approximations of Refs.~\onlinecite{ChaHea-JCP-08,LanRohHer-JPCB-08,RohHer-JCP-08,RohMarHer-JCP-09,PevTru-JPCL-11,NguDayPac-JCP-11,RefShaGovAutNeaBaeKro-PRL-12,EggWeiRefShaDauBaeKumNeaZojKro-JCTC-14,LufRefPacResRamKroPus-PRB-14,RefJaiShaNeaKro-PRB-15,LiuEggRefKroNea-JCP-17,BruProKroBre-JCP-17}, except that the part of correlation associated with the interaction $w_\ee^{\lr,\mu}(r_{12})+\l w_\ee^{\sr,\mu}(r_{12})$ is missing in Eq.~(\ref{Emul0}). The addition of this correlation is done in a second step with MP2 perturbation theory.

\subsection{Second-order M{\o}ller--Plesset perturbation theory}

A rigorous non-linear Rayleigh-Schr\"odinger perturbation theory starting from the single-determinant reference of Section~\ref{sec:sd} can be developed, similarly to what was done for the RSH+MP2 method in Refs.~\onlinecite{AngGerSavTou-PRA-05,FroJen-PRA-08,Ang-PRA-08} and for the one-parameter DH approximations in Ref.~\onlinecite{ShaTouSav-JCP-11}. This is done by introducing a perturbation strength parameter $\epsilon$ and defining the energy expression:
\begin{eqnarray}
E^{\mu,\l,\epsilon} = \min_{\Psi \rightarrow N} \Bigl\{ \bra{\Psi} \hat{T} + \hat{V}_\text{ne} + \hat{V}_{\Hx,\HF}^{\lr,\mu}[\Phi^{\mu,\l}] 
\nonumber\\
+ \l \hat{V}_{\Hx,\HF}^{\sr,\mu}[\Phi^{\mu,\l}] + \epsilon \hat{{\cal W}}^{\mu,\l}
\ket{\Psi} + \bar{E}_\Hxc^{\sr,\mu,\l}[n_\Psi] \Bigl\},
\label{GS-energy3}
\end{eqnarray}
where the search is over $N$-electron normalized antisymmetric multideterminant wave functions, and $\hat{{\cal W}}^{\mu,\l}$ is a M{\o}ller--Plesset-type perturbation operator
\begin{eqnarray}
\hat{{\cal W}}^{\mu,\l} &=& \hat{W}_\ee^{\lr,\mu} + \l \hat{W}_\ee^{\sr,\mu} 
\nonumber\\
&&- \hat{V}_{\Hx,\HF}^{\lr,\mu}[\Phi^{\mu,\l}] - \l \hat{V}_{\Hx,\HF}^{\sr,\mu}[\Phi^{\mu,\l}].
\end{eqnarray}
The minimizing wave function $\Psi^{\mu,\l,\epsilon}$ in Eq.~(\ref{GS-energy3}) is given by the corresponding Euler-Lagrange equation:
\begin{eqnarray}
\Bigl(\hat{T} + \hat{V}_{\text{ne}} + \hat{V}_{\Hx,\HF}^{\lr,\mu}[\Phi^{\mu,\l}] + \l \hat{V}_{\Hx,\HF}^{\sr,\mu}[\Phi^{\mu,\l}] 
+ \epsilon \hat{{\cal W}}^{\mu,\l}
\nonumber\\
+ \hat{V}_\Hxc^{\sr,\mu,\l}[n_{\Psi^{\mu,\l,\epsilon}}] \Bigl) \ket{\Psi^{\mu,\l,\epsilon}}
 = {\cal E}^{\mu,\l,\epsilon} \ket{\Psi^{\mu,\l,\epsilon}}. \;\;\;\;
\label{EL3}
\end{eqnarray}
For $\epsilon = 0$, Eq.~(\ref{EL3}) reduces to the single-determinant reference of Eq.~(\ref{EL2}), i.e.  $\Psi^{\mu,\l,\epsilon = 0} = \Phi^{\mu,\l}$ and ${\cal E}^{\mu,\l,\epsilon = 0} = {\cal E}_0^{\mu,\l}$. For $\epsilon = 1$, Eq.~(\ref{EL3}) reduces to Eq.~(\ref{EL1}), i.e. $\Psi^{\mu,\l,\epsilon = 1} = \Psi^{\mu,\l}$ and ${\cal E}^{\mu,\l,\epsilon = 1} = {\cal E}^{\mu,\l}$, and Eq.~(\ref{GS-energy3}) reduces to Eq.~(\ref{GS-energy1}), i.e. we recover the physical energy $E^{\mu,\l,\epsilon = 1} = E$, independently of $\mu$ and $\l$.
The perturbation theory is then obtained by expanding these quantities in $\epsilon$ around $\epsilon=0$: ${\cal E}^{\mu,\l,\epsilon}=\sum_{k=0}^{\infty} \epsilon^{k} {\cal E}^{\mu,\l,(k)}$, $\Psi^{\mu,\l,\epsilon}=\sum_{k=0}^{\infty} \epsilon^{k} \Psi^{\mu,\l,(k)}$, and $E^{\mu,\l,\epsilon}=\sum_{k=0}^{\infty} \epsilon^{k} E^{\mu,\l,(k)}$. Following the same steps as in Ref.~\onlinecite{AngGerSavTou-PRA-05}, we find the zeroth-order energy, 
\begin{eqnarray}
E^{\mu,\l,(0)} = \bra{\Phi^{\mu,\l}}  \hat{T} + \hat{V}_\text{ne} + \hat{V}_{\Hx,\HF}^{\lr,\mu}[\Phi^{\mu,\l}] 
\nonumber\\
+ \l \hat{V}_{\Hx,\HF}^{\sr,\mu}[\Phi^{\mu,\l}]  
\ket{\Phi^{\mu,\l}} + \bar{E}_\Hxc^{\sr,\mu,\l}[n_{\Phi^{\mu,\l}}],
\end{eqnarray}
and the first-order energy correction,
\begin{eqnarray}
E^{\mu,\l,(1)} = \bra{\Phi^{\mu,\l}} \hat{{\cal W}}^{\mu,\l} \ket{\Phi^{\mu,\l}},
\end{eqnarray}
so that the zeroth+first order energy gives back the energy of the single-determinant reference in Eq.~(\ref{Emul0}),
\begin{eqnarray}
E^{\mu,\l,(0)} + E^{\mu,\l,(1)} = E_0^{\mu,\l}.
\end{eqnarray}
The second-order energy correction involves only double-excited determinants $\Phi_{ij\to ab}^{\mu,\l}$ (of energy ${\cal E}_{0,ij\to ab}^{\mu,\l}$) and takes the form a MP2-like correlation energy, assuming a non-degenerate ground state in Eq.~(\ref{EL2}),
\begin{widetext}
\begin{eqnarray}
E^{\mu,\l,(2)} = E^{\mu,\l}_{\c,\MP} &=& - \sum_{i<j}^{\text{occ}} \sum_{a<b}^{\text{vir}} \frac{ \left|\bra{\Phi_{ij\to ab}^{\mu,\l}} \hat{{\cal W}}^{\mu,\l} \ket{\Phi^{\mu,\l}} \right|^2}{{\cal E}_{0,ij\to ab}^{\mu,\l} - {\cal E}_{0}^{\mu,\l}}
\nonumber\\
 &=& -\sum_{i<j}^{\text{occ}} \sum_{a<b}^{\text{vir}} \frac{\left|\bra{ij}\hat{w}_\ee^{\lr,\mu}+\l\hat{w}_\ee^{\sr,\mu} \ket{ab}- \bra{ij}\hat{w}_\ee^{\lr,\mu}+\l\hat{w}_\ee^{\sr,\mu} \ket{ba}\right|^2}{\varepsilon_{a} +\varepsilon_{b} -\varepsilon_{i} -\varepsilon_{j}},
\label{MP2}
\end{eqnarray}
\end{widetext}
where $i$ and $j$ refer to occupied spin orbitals and $a$ and $b$ refer to virtual spin orbitals obtained from Eq.~(\ref{EL2}), $\varepsilon_{k}$ are the associated orbital energies, and $\bra{ij}\hat{w}_\ee^{\lr,\mu}+\l\hat{w}_\ee^{\sr,\mu} \ket{ab}$ are the two-electron integrals corresponding to the interaction $w_\ee^{\lr,\mu}(r_{12})+\l w_\ee^{\sr,\mu}(r_{12})$. Note that the orbitals and orbital energies implicitly depend on $\mu$ and $\l$. Just like in standard M{\o}ller--Plesset perturbation theory, there is a Brillouin theorem making the single-excitation term vanish (see Ref.~\onlinecite{AngGerSavTou-PRA-05}). Also, contrary to the approach of Refs.~\onlinecite{CorStoJenFro-PRA-13,CorFro-IJQC-14}, the second-order energy correction does not involve any contribution from the second-order correction to the density.
The total RSDH energy is finally
\begin{eqnarray}
E^{\mu,\l}_{\text{RSDH}} = E^{\mu,\l}_0 + E^{\mu,\l}_{\c,\MP}.
\label{EmulRSDH}
\end{eqnarray}

It is instructive to decompose the correlation energy in Eq.~(\ref{MP2}) as
\begin{eqnarray}
E^{\mu,\l}_{\c,\MP} &=& E^{\lr,\mu}_{\c,\MP} + \l E^{\lr-\sr,\mu}_{\c,\MP} +  \l^2 E^{\sr,\mu}_{\c,\MP}, \;\;
\label{Emul2decomp}
\end{eqnarray}
with a pure long-range contribution,
\begin{eqnarray}
E_{\c,\MP}^{\lr,\mu} &=& - \sum_{i<j}^{\text{occ}} \sum_{a<b}^{\text{vir}} \frac{\left|\bra{ij}\hat{w}_\ee^{\lr,\mu}\ket{ab}- \bra{ij}\hat{w}_\ee^{\lr,\mu}\ket{ba}\right|^2}{\varepsilon_{a} +\varepsilon_{b} -\varepsilon_{i} -\varepsilon_{j}},
\nonumber\\
\end{eqnarray}
a pure short-range contribution,
\begin{eqnarray}
E_{\c,\MP}^{\sr,\mu} &=& -\sum_{i<j}^{\text{occ}} \sum_{a<b}^{\text{vir}} \frac{\left|\bra{ij}\hat{w}_\ee^{\sr,\mu}\ket{ab}- \bra{ij}\hat{w}_\ee^{\sr,\mu}\ket{ba}\right|^2}{\varepsilon_{a} +\varepsilon_{b} -\varepsilon_{i} -\varepsilon_{j}},
\nonumber\\
\end{eqnarray}
and a mixed long-range/short-range contribution,
\begin{widetext}
\begin{eqnarray}
E_{\c,\MP}^{\lr-\sr,\mu} &=& - \sum_{i<j}^{\text{occ}} \sum_{a<b}^{\text{vir}}  \frac{\left(\bra{ij}\hat{w}_\ee^{\lr,\mu}\ket{ab}- \bra{ij}\hat{w}_\ee^{\lr,\mu}\ket{ba}\right) \left(\bra{ab}\hat{w}_\ee^{\sr,\mu}\ket{ij}- \bra{ba}\hat{w}_\ee^{\sr,\mu}\ket{ij}\right)}{\varepsilon_{a} +\varepsilon_{b} -\varepsilon_{i} -\varepsilon_{j}} + \text{c.c.},
\label{sr-lr-MP2}
\end{eqnarray}
\end{widetext}
where c.c. stands for the complex conjugate. The exchange-correlation energy in the RSDH approximation is thus
\begin{eqnarray}
E^{\mu,\l}_{\xc,\text{RSDH}} = E_{\x,\HF}^{\lr,\mu} + \l E_{\x,\HF}^{\sr,\mu} + E_{\c,\MP}^{\lr,\mu}
\nonumber\\
+ \l E_{\c,\MP}^{\lr-\sr,\mu} + \l^2 E_{\c,\MP}^{\sr,\mu} + \bar{E}_{\xc}^{\sr,\mu,\l}[n]. \;\;
\end{eqnarray}
It remains to develop approximations for the complement short-range exchange-correlation density functional $\bar{E}_{\xc}^{\sr,\mu,\l}[n]$, which is done in Section~\ref{sec:srdf}.

\subsection{Complement short-range exchange-correlation density functional}
\label{sec:srdf}

\subsubsection{Decomposition into exchange and correlation}

The complement short-range exchange-correlation density functional $\bar{E}_\xc^{\sr,\mu,\l}[n]$ can be decomposed into exchange and correlation contributions,
\begin{equation}
\bar{E}_\xc^{\sr,\mu,\l}[n] = E_\x^{\sr,\mu,\l}[n] + \bar{E}_\c^{\sr,\mu,\l}[n],
\end{equation}
where the exchange part is defined with the KS single determinant $\Phi[n]$ and is linear with respect to $\l$,
\begin{eqnarray}
E_\x^{\sr,\mu,\l}[n] &=& \bra{\Phi[n]} (1-\l) \hat{W}_\ee^{\sr,\mu} \ket{\Phi[n]} - E_\H^{\sr,\mu,\l}[n]
\nonumber\\
&=& (1-\l) E_\x^{\sr,\mu}[n],
\label{Exsrmul}
\end{eqnarray}
where $E_\x^{\sr,\mu}[n]=E_\x^{\sr,\mu,\l=0}[n]$ is the usual short-range exchange density functional, as already introduced, e.g., in Ref.~\onlinecite{TouColSav-PRA-04}). Several (semi-)local approximations have been proposed for $E_\x^{\sr,\mu}[n]$ (see, e.g., Refs.~\onlinecite{Sav-INC-96,GilAdaPop-MP-96,IikTsuYanHir-JCP-01,HeyScuErn-JCP-03,TouSavFla-IJQC-04,TouColSav-PRA-04,TouColSav-JCP-05,GolWerSto-PCCP-05,GolWerStoLeiGorSav-CP-06,GolErnMoeSto-JCP-09}). By contrast, the complement short-range correlation density functional $\bar{E}_\c^{\sr,\mu,\l}[n]$ cannot be exactly expressed in terms of the short-range correlation density functional $\bar{E}_\c^{\sr,\mu}[n] = \bar{E}_\c^{\sr,\mu,\l=0}[n]$ of Ref.~\onlinecite{TouColSav-PRA-04} for which several (semi-)local approximations have been proposed~\cite{TouSavFla-IJQC-04,TouColSav-PRA-04,TouColSav-JCP-05,GolWerSto-PCCP-05,PazMorGorBac-PRB-06,GolWerStoLeiGorSav-CP-06,GolErnMoeSto-JCP-09}. Note that in the approach of Ref.~\onlinecite{CorFro-IJQC-14} the complement density functional was defined using the pure long-range interacting wave function $\Psi^{\mu} = \Psi^{\mu,\lambda=1}$ and it was possible, using uniform coordinate scaling relations, to find an exact expression for it in terms of previously studied density functionals. This is not the case in the present approach because the complement density functional is defined using the wave function $\Psi^{\mu,\lambda}$ obtained with both long-range and short-range interactions. As explained in the Appendix, uniform coordinate scaling relations do not allow one to obtain an exact expression for $\bar{E}_\c^{\sr,\mu,\l}[n]$ in terms of previously studied density functionals. Therefore, the difficulty lies in developing approximations for $\bar{E}_\c^{\sr,\mu,\l}[n]$. For this, we first give the exact expression of $\bar{E}_\c^{\sr,\mu,\l}[n]$ in the Coulomb limit $\mu\to 0$ (and the related high-density limit) and in the short-range limit $\mu\to\infty$ (and the related low-density limit).

\subsubsection{Expression of $\bar{E}_\c^{\sr,\mu,\l}[n]$ in the Coulomb limit $\mu \to 0$ and in the high-density limit}

The complement short-range correlation density functional $\bar{E}_\c^{\sr,\mu,\l}[n]$ can be written as
\begin{eqnarray}
\bar{E}_\c^{\sr,\mu,\l}[n] = E_\c[n] - E_\c^{\mu,\l}[n],
\label{barEcsrml}
\end{eqnarray}
where $E_\c[n]$ is the standard KS correlation density functional and $E_\c^{\mu,\l}[n]$ is the correlation density functional associated with the interaction $w_\ee^{\lr,\mu}(r_{12})+\l w_\ee^{\sr,\mu}(r_{12})$.

For $\mu=0$, the density functional $E_\c^{\mu=0,\l}[n] = E_\c^{\l}[n]$ corresponds to the correlation functional associated with the scaled Coulomb interaction $\l w_\ee(r_{12})$, which can be exactly expressed as $E_\c^{\l}[n]=\l^2 E_\c[n_{1/\l}]$~\cite{Lev-PRA-91,Lev-INC-95} where $n_{1/\l}(\b{r}) = (1/\l^3) n(\b{r}/\l)$ is the density with coordinates uniformly scaled by $1/\l$. Therefore, for $\mu=0$, the complement short-range correlation density functional is
\begin{eqnarray}
\bar{E}_\c^{\sr,\mu=0,\l}[n] = E_\c[n] - \l^2 E_\c[n_{1/\l}],
\label{barEcl}
\end{eqnarray}
which is the correlation functional used in the density-scaled one-parameter double-hybrid (DS1DH) scheme of Sharkas \textit{et al.}~\cite{ShaTouSav-JCP-11}. For a KS system with a non-degenerate ground state, we have in the $\l\to 0$ limit: $E_\c[n_{1/\l}] = E_\c^{\text{GL2}}[n] + {\cal O}(\l)$ where $E_\c^{\text{GL2}}[n]$ is the second-order G\"orling--Levy (GL2) correlation energy~\cite{GorLev-PRB-93}. Therefore, in this case, $\bar{E}_\c^{\sr,\mu=0,\l}[n]$ has a quadratic dependence in $\l$ near $\l=0$. In practice with GGA functionals, it has been found that the density scaling in Eq.~(\ref{barEcl}) can sometimes be advantageously neglected, i.e. $E_\c[n_{1/\l}] \approx E_\c[n]$~\cite{ShaTouSav-JCP-11,ShaSavJenTou-JCP-12}, giving
\begin{eqnarray}
\bar{E}_\c^{\sr,\mu=0,\l}[n] \approx (1-\l^2) E_\c[n].
\label{barEclapp}
\end{eqnarray}

Even if we do not plan to apply the RSDH scheme with $\mu=0$, the condition in Eq.~(\ref{barEcl}) is in fact relevant for an arbitrary value of $\mu$ in the high-density limit, i.e. $n_\gamma(\b{r})= \gamma^3 n(\gamma \b{r})$ with $\gamma\to\infty$, since in this limit the short-range interaction becomes equivalent to the Coulomb interaction in the complement short-range correlation density functional: $\lim_{\gamma\to\infty} \bar{E}_\c^{\sr,\mu,\l}[n_\gamma] = \lim_{\gamma\to\infty} \bar{E}_\c^{\sr,\mu=0,\l}[n_\gamma]$ (see Appendix). In fact, for a KS system with a non-degenerate ground state, the approximate condition in Eq.~(\ref{barEclapp}) is sufficient to recover the exact high-density limit for an arbitrary value of $\mu$ which is (see Appendix)
\begin{equation}
\lim_{\gamma\to\infty} \bar{E}_\c^{\sr,\mu,\l}[n_\gamma] = (1-\l^2) E_\c^{\text{GL2}}[n].
\end{equation}

\subsubsection{Expression of $\bar{E}_\c^{\sr,\mu,\l}[n]$ in the short-range limit $\mu\to\infty$ and the low-density limit}

The leading term in the asymptotic expansion of $\bar{E}_\c^{\sr,\mu,\l}[n]$ as $\mu\to\infty$ is (see Appendix)
\begin{eqnarray}
\bar{E}_\c^{\sr,\mu,\l}[n]&=& (1-\l) \frac{\pi}{2\mu^2}  \int n_{2,\c}[n](\b{r},\b{r}) \d\b{r} 
\nonumber\\
&&\;\;\;\;+ \; {\cal O}\left( \frac{1}{\mu^3}\right),
\label{Ecmulmuinfty}
\end{eqnarray}
where $n_{2,\c}[n](\b{r},\b{r})$ is the correlation part of the on-top pair density for the Coulomb interaction. We thus see that, for $\mu\to\infty$, $\bar{E}_\c^{\sr,\mu,\l}[n]$ is linear with respect to $\l$. In fact, since the asymptotic expansion of the usual short-range correlation functional is $\bar{E}_\c^{\sr,\mu}[n] = \pi/(2\mu^2) \int n_{2,\c}[n](\b{r},\b{r}) \d\b{r} + {\cal O}(1/\mu^3)$~\cite{TouColSav-PRA-04}, we can write for $\mu\to\infty$,
\begin{eqnarray}
\bar{E}_\c^{\sr,\mu,\l}[n] = (1-\l) \bar{E}_\c^{\sr,\mu}[n]  + {\cal O}\left( \frac{1}{\mu^3}\right).
\label{EcmulmuinftyEcmu}
\end{eqnarray}

The low-density limit, i.e. $n_\gamma(\b{r})= \gamma^3 n(\gamma \b{r})$ with $\gamma\to 0$, is closely related to the limit $\mu\to \infty$ (see Appendix)
\begin{eqnarray}
\bar{E}_\c^{\sr,\mu,\lambda}[n_\gamma]
&\isEquivTo{\gamma \to 0}& \frac{ \gamma^3 (1-\l)\pi}{2\mu^2}  \int n_{2,\c}^{1/\gamma}[n](\b{r},\b{r}) \d\b{r} 
\nonumber\\
&\isEquivTo{\gamma \to 0}& \frac{\gamma^3 (1-\l)\pi}{4\mu^2}  \int \!\! \left[ -n(\b{r})^2 + m(\b{r})^2\right] \! \d\b{r}, 
\nonumber\\
\label{Ecmulgamma0}
\end{eqnarray}
in which appears $n_{2,\c}^{1/\gamma}[n](\b{r},\b{r})$, the on-top pair density for the scaled Coulomb interaction $(1/\gamma) w_\ee(r_{12})$, and its strong-interaction limit $\lim_{\gamma\to 0}n_{2,\c}^{1/\gamma}[n](\b{r},\b{r}) = -n(\b{r})^2/2 + m(\b{r})^2/2$~\cite{BurPerErn-JCP-98} where $m(\b{r})$ is the spin magnetization. Thus, in the low-density limit, contrary to the usual KS correlation functional $E_\c[n]$ which goes to zero linearly in $\gamma$~\cite{Lev-PRA-91}, times the complicated nonlocal strictly-correlated electron functional~\cite{SeiGorSav-PRA-07}, the complement short-range correlation functional $\bar{E}_\c^{\sr,\mu,\lambda}[n]$ goes to zero like $\gamma^3$ and becomes a simple local functional of $n(\b{r})$ and $m(\b{r})$.

\subsubsection{Approximations for $\bar{E}_\c^{\sr,\mu,\l}[n]$}
\label{sec:approx}

We now propose several simple approximations for $\bar{E}_\c^{\sr,\mu,\l}[n]$. On the one hand, Eq.~(\ref{barEclapp}) suggests the approximation
\begin{eqnarray}
\bar{E}_{\c,\text{approx1}}^{\sr,\mu,\l}[n] = (1-\l^2) \bar{E}_\c^{\sr,\mu}[n], 
\label{approx.1}
\end{eqnarray}
which is correctly quadratic in $\l$ at $\mu=0$ but is not linear in $\l$ for $\mu\to\infty$. On the other hand, Eq.~(\ref{EcmulmuinftyEcmu}) suggests the approximation
\begin{eqnarray}
\bar{E}_{\c,\text{approx2}}^{\sr,\mu,\l}[n] = (1-\l) \bar{E}_\c^{\sr,\mu}[n],
\label{approx.2} 
\end{eqnarray}
which is correctly linear in $\l$ for $\mu\to\infty$ but not quadratic in $\l$ at $\mu=0$. 

However, it is possible to impose simultaneously the two limiting behaviors for $\mu=0$ and $\mu\to\infty$ with the following approximation
\begin{eqnarray}
\bar{E}_{\c,\text{approx3}}^{\sr,\mu,\l}[n] = \bar{E}_\c^{\sr,\mu}[n] - \l^2 \bar{E}_\c^{\sr,\mu\sqrt{\l}}[n],
\label{approx.3}
\end{eqnarray}
which reduces to Eq.~(\ref{barEclapp}) for $\mu=0$ and satisfies Eq.~(\ref{Ecmulmuinfty}) for $\mu\to\infty$. Another possibility, proposed in Ref.~\onlinecite{GarBulHenScu-PCCP-15}, is
\begin{eqnarray}
\bar{E}_{\c,\text{approx4}}^{\sr,\mu,\l}[n] = \bar{E}_\c^{\sr,\mu}[n] - \l^2 \bar{E}_\c^{\sr,\mu/\l}[n_{1/\l}],
\label{approx.4}
\end{eqnarray}
which correctly reduces to Eq.~(\ref{barEcl}) for $\mu=0$. For $\mu\to\infty$, its asymptotic expansion is
\begin{eqnarray}
\bar{E}_{\c,\text{approx4}}^{\sr,\mu,\l}[n]= \frac{\pi}{2\mu^2}  \int n_{2,\c}[n](\b{r},\b{r}) \d\b{r} \phantom{xxxx}
\nonumber\\
- \l^4 \frac{\pi}{2\mu^2}  \int n_{2,\c}[n_{1/\l}](\b{r},\b{r}) \d\b{r} + {\cal O}\left( \frac{1}{\mu^3}\right),
\label{Ecmulmuinftyapprox.4}
\end{eqnarray}
i.e. it does not satisfy Eq.~(\ref{Ecmulmuinfty}). Contrary to what was suggested in Ref.~\onlinecite{GarBulHenScu-PCCP-15}, Eq.~(\ref{approx.4}) is not exact but only an approximation. However, using the scaling relation on the system-averaged on-top pair density~\cite{BurPerErn-JCP-98}
\begin{eqnarray}
\int n_{2,\c}[n_{\gamma}](\b{r},\b{r}) \d\b{r}= \gamma^3 \!\! \int n_{2,\c}^{1/\gamma}[n](\b{r},\b{r}) \d\b{r}, \;\;
\end{eqnarray}
it can be seen that, in the low-density limit $\gamma\to 0$, Eq.~(\ref{Ecmulmuinftyapprox.4}) correctly reduces to Eq.~(\ref{Ecmulgamma0}). In Ref.~\onlinecite{GarBulHenScu-PCCP-15}, the authors propose to neglect the scaling of the density in Eq.~(\ref{approx.4}) leading to
\begin{eqnarray}
\bar{E}_{\c,\text{approx5}}^{\sr,\mu,\l}[n] = \bar{E}_\c^{\sr,\mu}[n] - \l^2 \bar{E}_\c^{\sr,\mu/\l}[n],
\label{approx.5}
\end{eqnarray}
which reduces to Eq.~(\ref{barEclapp}) for $\mu=0$, but which has also a wrong $\lambda$-dependence for large $\mu$
\begin{eqnarray}
\bar{E}_{\c,\text{approx5}}^{\sr,\mu,\l}[n]&=& (1- \l^4 )\frac{\pi}{2\mu^2}  \int n_{2,\c}[n](\b{r},\b{r}) \d\b{r} 
\nonumber\\
&&\;\;\;\;\;\;\;+ {\cal O}\left( \frac{1}{\mu^3}\right),
\end{eqnarray}
and does not anymore satisfy the low-density limit.

Another strategy is to start from the decomposition of the MP2-like correlation energy in Eq.~(\ref{Emul2decomp}) which suggests the following approximation for the complement short-range correlation functional
\begin{eqnarray}
\bar{E}_{\c,\text{approx6}}^{\sr,\mu,\l}[n]&=& (1-\l) E_{\c}^{\lr-\sr,\mu}[n] 
\nonumber\\
&&+ (1-\l^2) E_{\c}^{\sr,\mu}[n],
\label{approx.6}
\end{eqnarray}
where $E_{\c}^{\lr-\sr,\mu}[n]=\bar{E}_{\c}^{\sr,\mu}[n]-E_{\c}^{\sr,\mu}[n]$ is the mixed long-range/short-range correlation functional~\cite{Tou-THESIS-05,TouSav-JMS-06} and $E_{\c}^{\sr,\mu}[n]$ is the pure short-range correlation functional associated with the short-range interaction $w_{\text{ee}}^{\sr,\mu}(r_{12})$~\cite{Tou-THESIS-05,TouSav-JMS-06}. An LDA functional has been constructed for $E_{\c}^{\sr,\mu}[n]$~\cite{ZecGorMorBac-PRB-04}. Since $E_{\c}^{\lr-\sr,\mu=0}[n]=0$ and $E_{\c}^{\sr,\mu=0}[n]=E_{\c}[n]$, the approximation in Eq.~(\ref{approx.6}) reduces to Eq.~(\ref{barEclapp}) for $\mu=0$. For $\mu\to\infty$, since $E_{\c}^{\sr,\mu}[n]$ decays faster than $1/\mu^2$, i.e. $E_{\c}^{\sr,\mu}[n] = {\cal O}( 1/\mu^3)$~\cite{ZecGorMorBac-PRB-04}, $E_{\c}^{\lr-\sr,\mu}[n]$  and $\bar{E}_{\c}^{\sr,\mu}[n]$ have the same leading term in the large-$\mu$ expansion, i.e. $E_{\c}^{\lr-\sr,\mu}[n]=\bar{E}_{\c}^{\sr,\mu}[n] + {\cal O}(1/\mu^3)$, and thus the approximation in Eq.~(\ref{approx.6}) satisfies Eq.~(\ref{Ecmulmuinfty}) or (\ref{EcmulmuinftyEcmu}). One can also enforce the exact condition at $\mu = 0$, Eq.~(\ref{barEclapp}), by introducing a scaling of the density
\begin{eqnarray}
\bar{E}_{\c,\text{approx7}}^{\sr,\mu,\l}[n]&=& (1-\l) E_{\c}^{\lr-\sr,\mu}[n] + E_{\c}^{\sr,\mu}[n]
\nonumber\\
&&-\l^2 E_{\c}^{\sr,\mu/\l}[n_{1/\l}].
\label{approx.7}
\end{eqnarray}

\subsubsection{Assessment of the approximations for $\bar{E}_\c^{\sr,\mu,\l}[n]$ on the uniform-electron gas}

We now test the approximations for the complement short-range correlation functional $\bar{E}_\c^{\sr,\mu,\l}[n]$ introduced in Sec.~\ref{sec:approx} on the spin-unpolarized uniform-electron gas. 

As a reference, for several values of the Wigner-Seitz radius $r_\s=(3/(4\pi n))^{1/3}$ and the parameters $\mu$ and $\l$, we have calculated the complement short-range correlation energy per particle as
\begin{eqnarray}
\bar{\varepsilon}_{\c,\text{unif}}^{\sr,\mu,\l}(r_\s) = \varepsilon_{\c,\text{unif}}(r_\s) - \varepsilon_{\c,\text{unif}}^{\mu,\l}(r_\s),
\end{eqnarray}
where $\varepsilon_{\c,\text{unif}}(r_\s)$ is the correlation energy per particle of the uniform-electron gas with the Coulomb electron-electron $w_\ee(r_{12})$ and $\varepsilon_{\c,\text{unif}}^{\mu,\l}(r_\s)$ is the correlation energy per particle of an uniform-electron gas with the modified electron-electron $w_\ee^{\lr,\mu}(r_{12})+\l w_\ee^{\sr,\mu}(r_{12})$. We used what is known today as the direct random-phase approximation + second-order screened exchange (dRPA+SOSEX) method (an approximation to coupled-cluster doubles)~\cite{TouZhuSavJanAng-JCP-11,GruMarHarSchKre-JCP-09} as introduced for the uniform-electron gas by Freeman~\cite{Fre-PRB-77} and extended for modified electron-electron interactions in Refs.~\onlinecite{Sav-INC-96,TouSavFla-IJQC-04}, and which is known to give reasonably accurate correlation energies per particle of the spin-unpolarized electron gas (error less than 1 millihartree for $r_\s < 10$). We note that these calculations would allow us to construct a complement short-range LDA correlation functional, but we refrain from doing that since we prefer to avoid having to do a complicated fit of $\bar{\varepsilon}_\c^{\sr,\mu,\l}(r_\s)$ with respect to $r_\s$, $\mu$, and $\l$. Moreover, this would only give a spin-independent LDA functional. We thus use these uniform-electron gas calculations only to test the approximations of Sec.~\ref{sec:approx}.
 
\begin{figure}[t]
\centering
   \includegraphics[scale=0.60,angle=0]{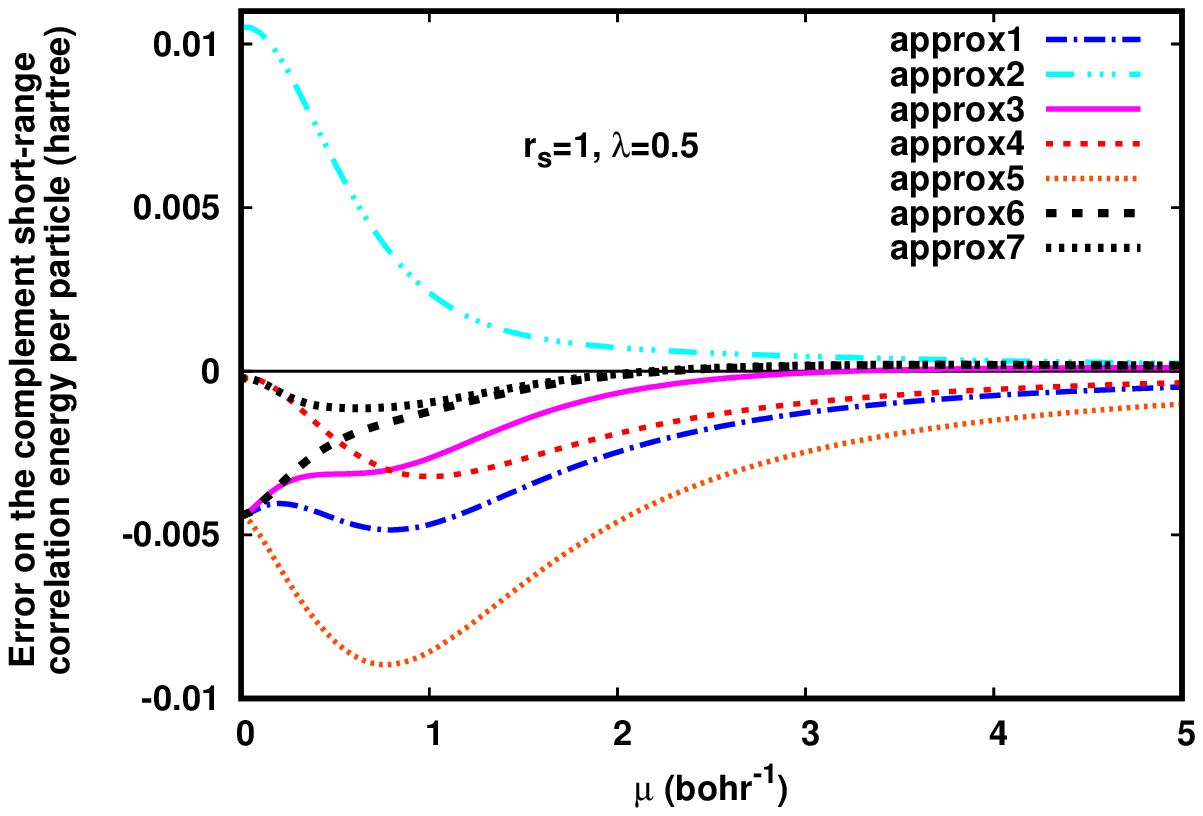} \\
   \includegraphics[scale=0.60,angle=0]{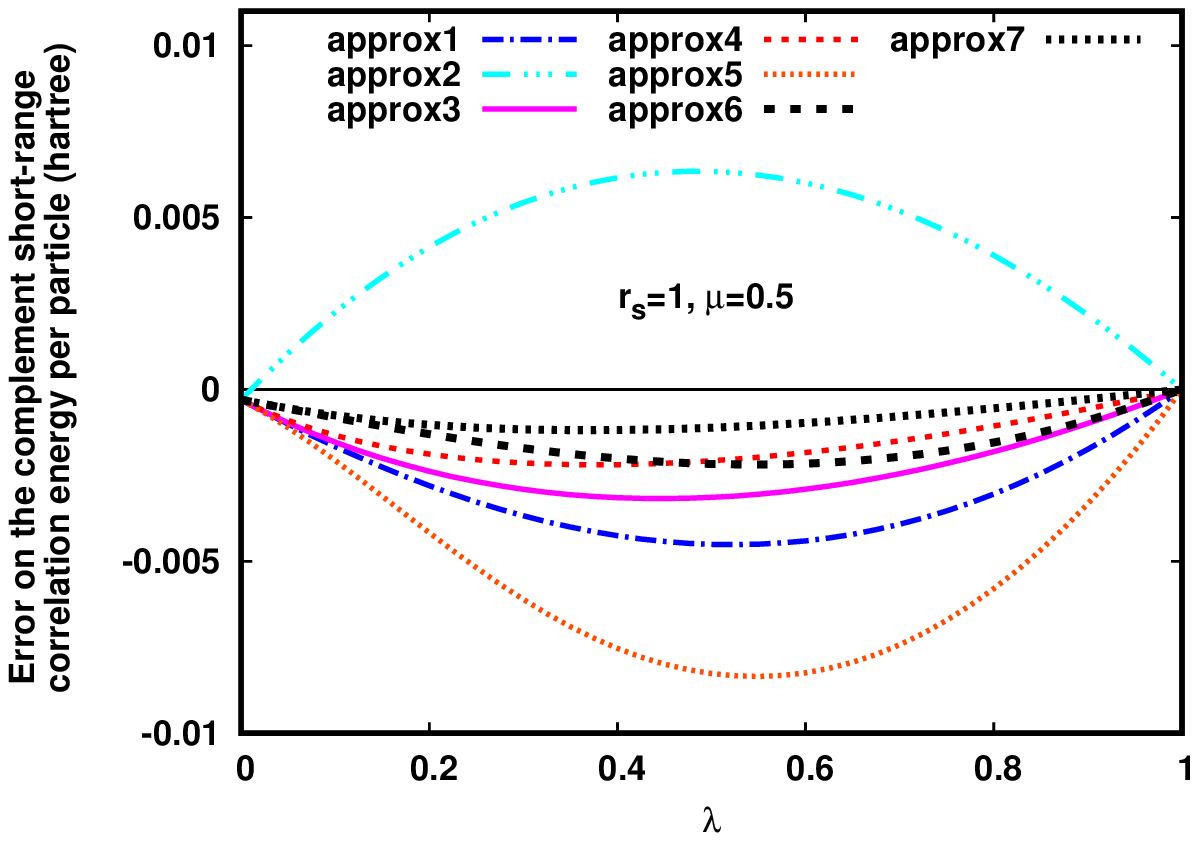} \\
   \includegraphics[scale=0.60,angle=0]{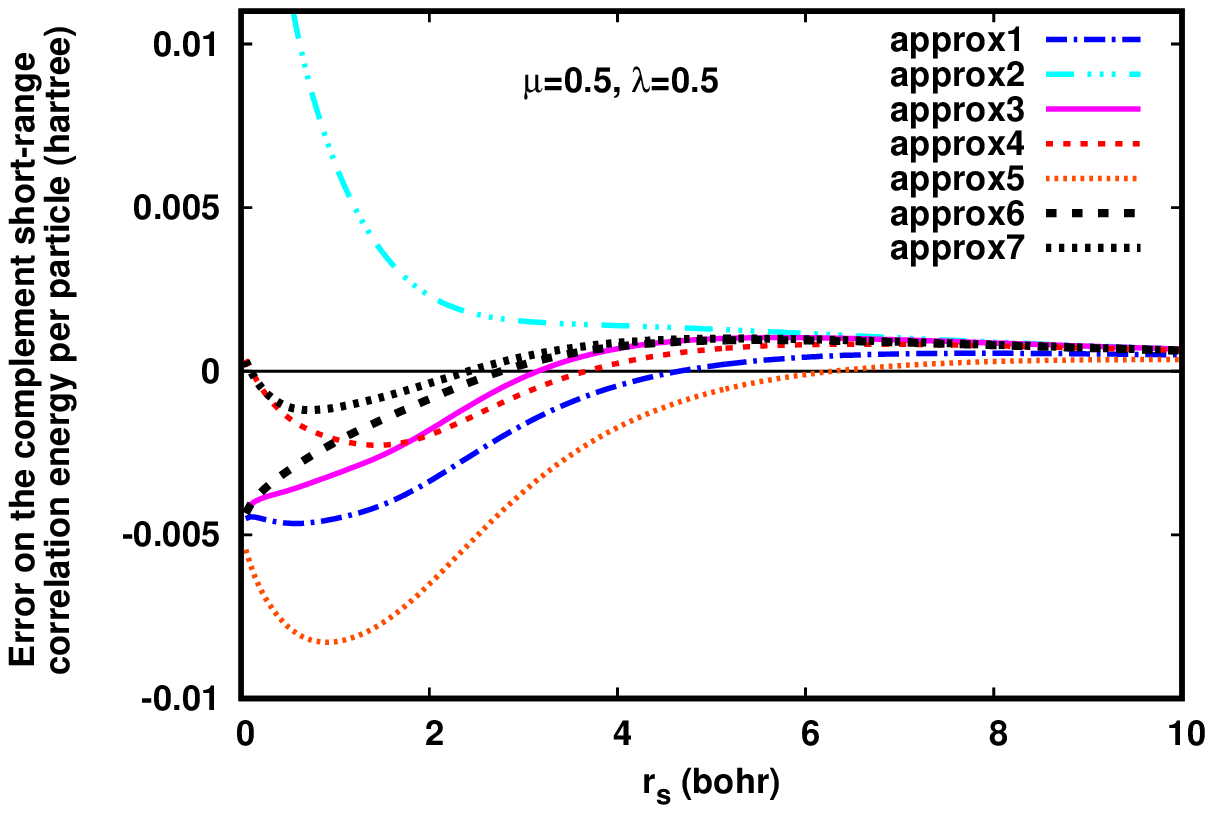}
   \caption{Error on the complement short-range correlation energy per particle $\bar{\varepsilon}_{\c,\text{unif}}^{\sr,\mu,\l}(r_\s)$ of the uniform-electron gas obtained with approximations 1 to 7 of Sec.~\ref{sec:approx} with respect to the dRPA+SOSEX results.} 
        \label{fig:ueg}
\end{figure}

For several values of $r_\s$, $\mu$, and $\l$, we have calculated the complement short-range correlation energy per particle corresponding to the approximations 1 to 7 using the LDA approximation for $\bar{E}_{\c}^{\sr,\mu}[n]$ from Ref.~\onlinecite{PazMorGorBac-PRB-06} (for approximations 1 to 7), as well as the LDA approximation for $E_{\c}^{\sr,\mu}[n]$ from Ref.~\onlinecite{ZecGorMorBac-PRB-04} (for approximations 6 and 7), and the errors with respect to the dRPA+SOSEX results are reported in Fig.~\ref{fig:ueg}. Note that the accuracy of the dRPA+SOSEX reference decreases as $r_\s$ increases, the error being of the order of 1 millihartree for $r_\s=10$, which explains why the curves on the third graph of Fig.~\ref{fig:ueg} appear shifted with respect to zero at large $r_\s$.

By construction, all the approximations become exact for $\l=0$ (and trivially for $\l=1$ or in the $\mu\to\infty$ limit since the complement short-range correlation energy goes to zero in these cases). For intermediate values of $\l$ and finite values of $\mu$, all the approximations, except approximation 2, tend to give too negative correlation energies. As it could have been expected, approximation 2, which is the only one incorrectly linear in $\l$ at $\mu=0$, gives quite a large error (of the order of 0.01 hartree or more) for small $\mu$, intermediate $\l$, and small $r_\s$ (it in fact diverges in the high-density limit $r_\s \to 0$), but the error goes rapidly to zero as $\mu$ increases, reflecting the fact that this approximation has the correct leading term of the asymptotic expansion for $\mu\to\infty$. On the contrary, approximation 1, being quadratic in $\l$, gives a smaller error (less than 0.005 hartree) for small $\mu$ but the error goes slower to zero as $\mu$ increases. Approximation 3 combines the advantages of approximations 1 and 2: it gives a small error for small $\mu$ which goes rapidly to zero as $\mu$ increases. Approximation 4, which contains the scaling of the density, is exact for $\mu=0$, and gives a small error (at most about 0.003 hartree) for intermediate values of $\mu$, but the error does not go rapidly to zero as $\mu$ increases. Again, this reflects the fact that this approximation does not give the correct leading term of the asymptotic expansion for $\mu\to\infty$ for arbitrary $\l$ and $r_\s$. This confirms that Eq.~(\ref{approx.4}) does not give the exact complement short-range correlation functional, contrary to what was thought in Ref.~\onlinecite{GarBulHenScu-PCCP-15}. A nice feature however of approximation 4 is that it becomes exact in the high-density limit $r_\s \to 0$ of the uniform-electron gas (the scaling of the density at $\mu=0$ is needed to obtain the correct high-density limit in this zero-gap system). Approximation 5, obtained from approximation 4 by neglecting the scaling of the density in the correlation functional, and used in Ref.~\onlinecite{GarBulHenScu-PCCP-15}, gives quite large errors for the uniform-electron gas, approaching 0.01 hartree. Approximations 6 and 7 are quite good. They both have the correct leading term of the asymptotic expansion for $\mu\to\infty$, but approximation 7 has the additional advantage of having also the correct $\mu\to0$ or $r_\s\to0$ limit. Approximation 7 is our best approximation, with a maximal error of about 1 millihartree.

Unfortunately, approximations 6 and 7 involve the pure short-range correlation functional $E_{\c}^{\sr,\mu}[n]$, for which we currently have only a spin-unpolarized LDA approximation~\cite{ZecGorMorBac-PRB-04}. For this reason, we do not consider these approximations in the following for molecular calculations. We will limit ourselves to approximations 1 to 5 which only involve the complement short-range correlation functional $\bar{E}_{\c}^{\sr,\mu}[n]$, for which we have spin-dependent GGAs~\cite{TouColSav-PRA-04,TouColSav-JCP-05,GolWerSto-PCCP-05,GolWerStoLeiGorSav-CP-06,GolErnMoeSto-JCP-09}.

\begin{figure*}[t]
\centering
\includegraphics[scale=0.50,angle=0]{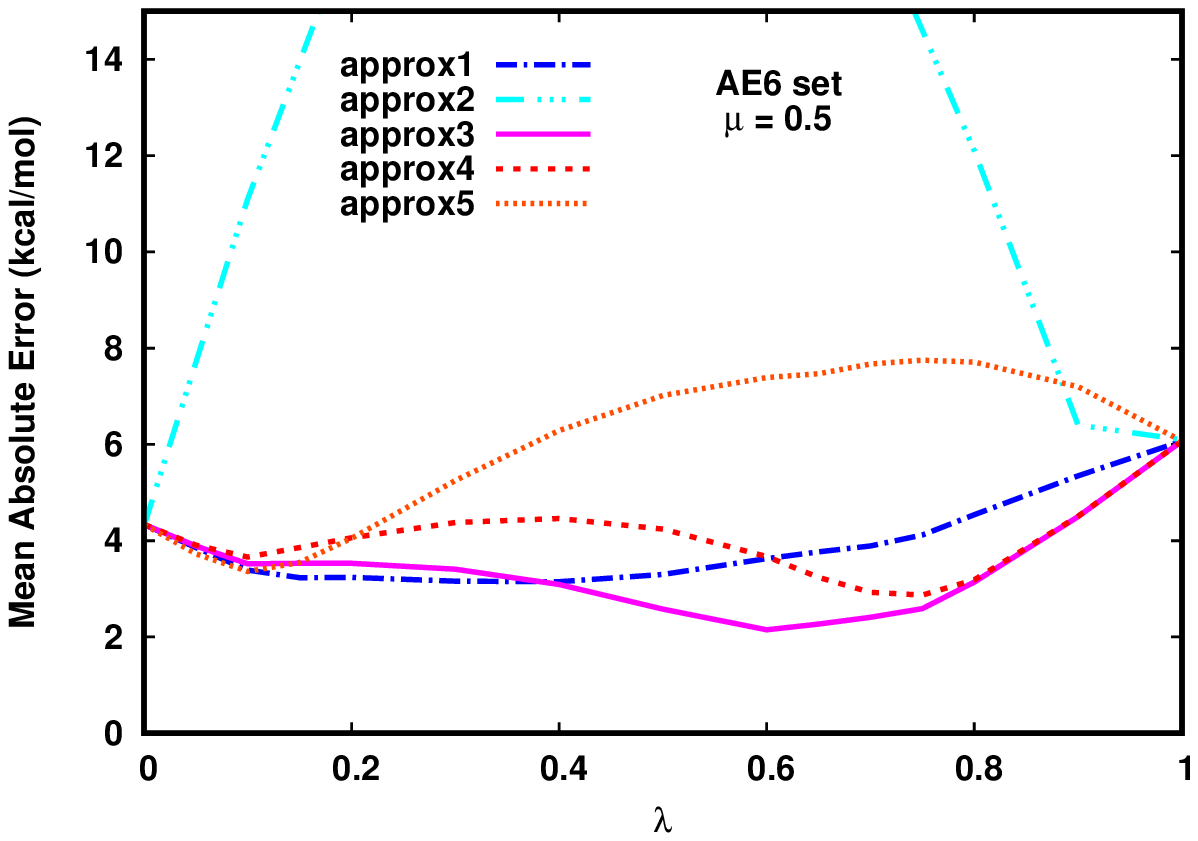}
\includegraphics[scale=0.50,angle=0]{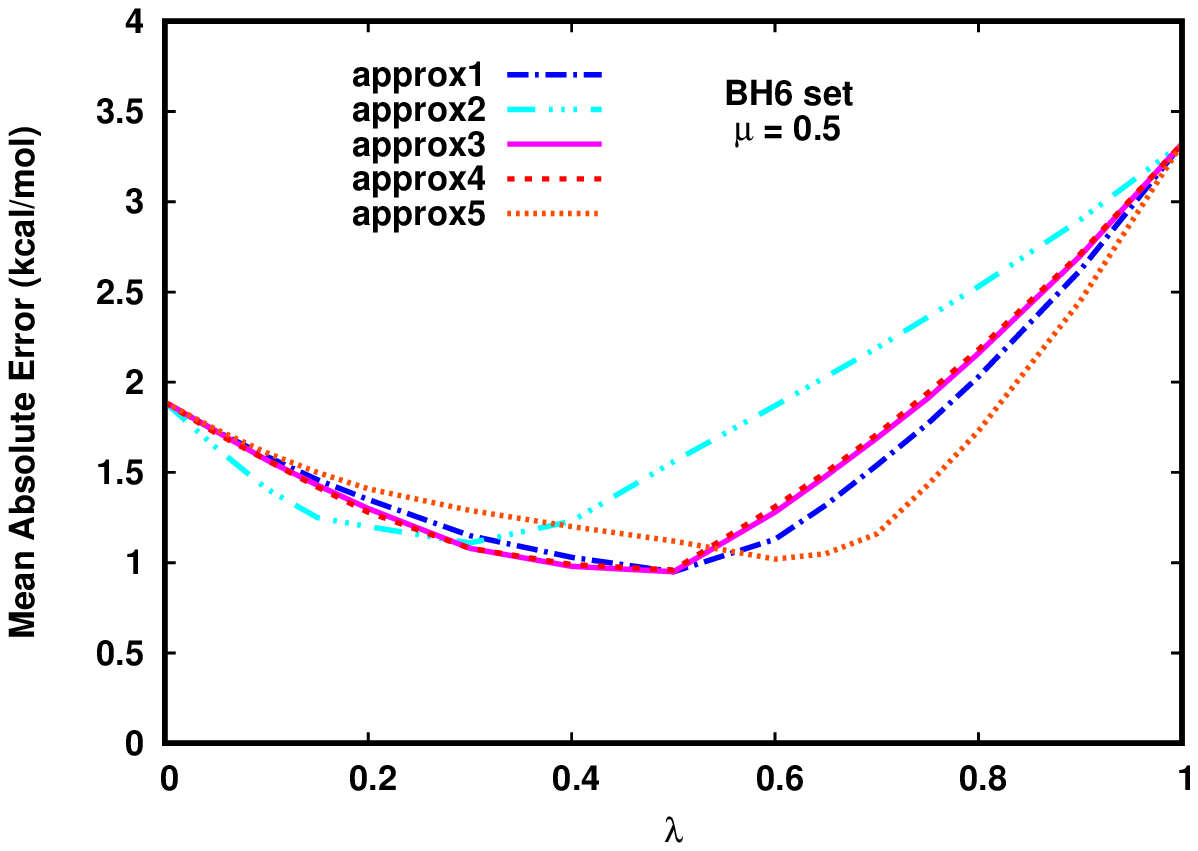}\\
\includegraphics[scale=0.50,angle=0]{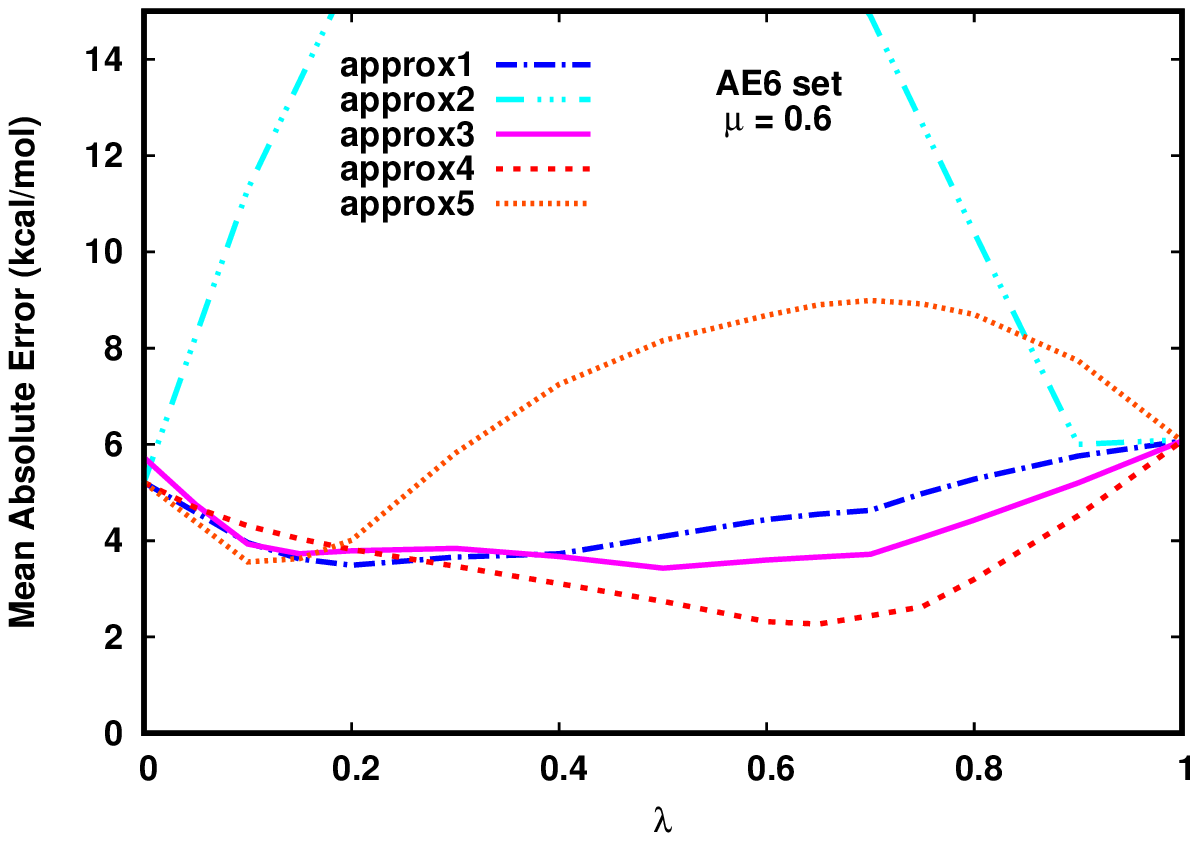}
\includegraphics[scale=0.50,angle=0]{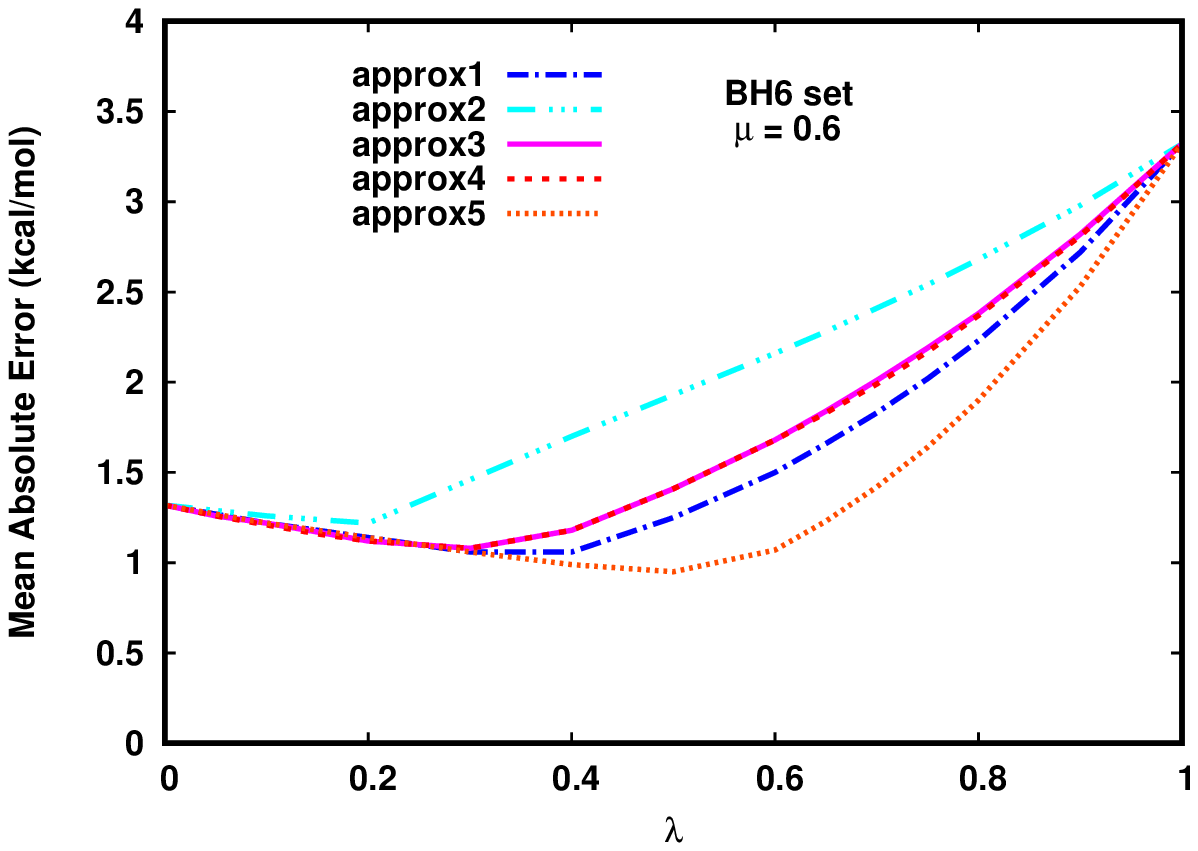}
\caption{MAEs for the AE6 and BH6 sets obtained with the RSDH scheme using approximations 1 to 5 of Sec.~\ref{sec:approx} (with the short-range PBE exchange-correlation functional of Ref.~\onlinecite{GolWerStoLeiGorSav-CP-06}) as a function of $\l$ for $\mu=0.5$ and $\mu=0.6$. The basis set used is cc-pVQZ.}
\label{AE6_BH6}
\end{figure*}

\section{Computational details}
\label{sec:comp}

The RSDH scheme has been implemented in a development version of the {\sc MOLPRO 2015} program~\cite{Molproshort-PROG-15}. The calculation is done in two steps: first a self-consistent-field calculation is perform according to Eqs.~(\ref{RSH-1})-(\ref{Emul0}), and then the MP2-like correlation energy in Eq.~(\ref{MP2}) is evaluated with the previously calculated orbitals. The $\l$-dependent complement short-range exchange functional is calculated according to Eq.~(\ref{Exsrmul}) and the approximations 1 to 5 [see Eqs.~(\ref{approx.1})-(\ref{approx.5})] have been implemented for the complement short-range correlation functional, using the short-range Perdew-Becke-Ernzerhof (PBE) exchange and correlation functionals of Ref.~\onlinecite{GolWerStoLeiGorSav-CP-06} for $E_\x^{\sr,\mu}[n]$ and $\bar{E}_\c^{\sr,\mu}[n]$. 

The RSDH scheme was applied on the AE6 and BH6 sets~\cite{LynTru-JPCA-03}, as a first assessment of the approximations on molecular systems and in order to determine the optimal parameters $\mu$ and $\l$. The AE6 set is a small representative benchmark of six atomization energies consisting of SiH$_4$, S$_2$, SiO, C$_3$H$_4$ (propyne), C$_2$H$_2$O$_2$ (glyoxal), and C$_4$H$_8$ (cyclobutane). The BH6 set is a small representative benchmark of forward and reverse hydrogen transfer barrier heights of three reactions, OH + CH$_4$ $\rightarrow$ CH$_3$ + H$_2$O, H + OH $\rightarrow$ O + H$_2$, and H + H$_2$S $\rightarrow$ HS + H$_2$. All the calculations for the AE6 and BH6 sets were performed with the Dunning cc-pVQZ basis set~\cite{Dun-JCP-89} at the geometries optimized by quadratic configuration interaction with single and double excitations with the modified Gaussian-3 basis set (QCISD/MG3)~\cite{KalTou-JJJ-XX-note1}. The reference values for the atomization energies and barrier heights are the non-relativistic FC-CCSD(T)/cc-pVQZ-F12 values of Refs.~\onlinecite{HauKlo-TCA-12,HauKlo-TCA-12-err}. For each approximation, we have first varied $\mu$ and $\l$ between 0 and 1 by steps of $0.1$ to optimize the parameters on each set. We have then refined the search by steps of 0.02 to find the common optimal parameters on the two sets combined.

The RSDH scheme was then tested on the AE49 set of 49 atomization energies~\cite{FasCorSanTru-JPCA-99} (consisting of the G2-1 set~\cite{CurRagTruPop-JCP-91,CurRagRedPop-JCP-97} stripped of the six molecules containing Li, Be, and Na~\cite{KalTou-JJJ-XX-note3}) and on the DBH24/08 set~\cite{ZheZhaTru-JCTC-07,ZheZhaTru-JCTC-09} of 24 forward and reverse reaction barrier heights. These calculations were performed with the cc-pVQZ basis set, with MP2(full)/6-31G* geometries for the AE49 set, and with the aug-cc-pVQZ basis set~\cite{WooDun-JCP-93} with QCISD/MG3 geometries for the DBH24/08 set. The reference values for the AE49 set are the non-relativistic FC-CCSD(T)/cc-pVQZ-F12 values of Ref.~\onlinecite{HauKlo-JCP-12}, and the reference values for the DBH24/08 set are the zero-point exclusive values from Ref.~\onlinecite{ZheZhaTru-JCTC-09}.

\begingroup
\squeezetable
\begin{table*}
\renewcommand{\arraystretch}{0.8}
\setlength{\tabcolsep}{0.07cm}
\footnotesize
\caption{Atomization energies (in kcal/mol) of the AE49 set calculated by DS1DH (with the PBE exchange-correlation functional~\cite{PerBurErn-PRL-96}), RSH+MP2, RSDH with approximations 3 and 4 of Sec.~\ref{sec:approx} (with the short-range PBE exchange-correlation functional of Ref.~\onlinecite{GolWerStoLeiGorSav-CP-06}), and MP2. The calculations were carried out using the cc-pVQZ basis set at MP2(full)/6-31G* geometries and with parameters $(\mu,\l)$ optimized on the AE6+BH6 combined set. The reference values are the non-relativistic FC-CCSD(T)/cc-pVQZ-F12 values of Ref.~\onlinecite{HauKlo-JCP-12}.}
\label{tab:AE49}
\begin{tabular}{lcccccc}
\hline\hline\\[-0.1cm]
Molecule                & DS1DH         &  \phantom{xx} RSH+MP2 \phantom{xx} & RSDH approx3   & RSDH approx4   & MP2                       & Reference    \\
\phantom{xxxxxxxx} $(\mu$,$\l)=$        & (0,0.70) & (0.58,0)  & (0.46,0.58) & (0.62,0.60) &                       &        \\
\hline\\[-0.1cm]
CH                      & 81.13  & 78.38   & 79.93     & 79.58     & 79.68                     & 83.87  \\
CH$_{2}$($^{3}$B$_{1})$ & 190.68 & 190.19  & 190.42    & 190.32    & 188.70                    & 189.74 \\
CH$_{2}$($^{1}$A$_{1})$ & 175.20 & 170.26  & 173.36    & 173.24    & 174.45                    & 180.62 \\
CH$_{3}$                & 305.32 & 302.91  & 304.43    & 304.34    & 303.36                    & 306.59 \\
CH$_{4}$                & 415.79 & 410.84  & 414.23    & 414.45    & 414.83                    & 418.87 \\
NH                      & 81.39  & 81.09   & 80.28     & 79.49     & 78.57                     & 82.79  \\
NH$_{2}$                & 179.12 & 177.12  & 177.03    & 176.22    & 176.65                    & 181.96 \\
NH$_{3}$                & 293.24 & 288.76  & 290.33    & 290.02    & 293.11                    & 297.07 \\
OH                      & 105.26 & 104.49  & 104.02    & 103.81    & 105.78                    & 106.96 \\
OH$_{2}$                & 229.78 & 225.48  & 226.96    & 227.21    & 233.83                    & 232.56 \\
FH                      & 140.11 & 137.20   & 138.16    & 138.35    & 144.17                    & 141.51 \\
SiH$_{2}$($^{1}$A$_{1})$ & 146.66 & 143.21  & 146.22    & 146.15    & 145.90                    & 153.68 \\
SiH$_{2}$($^{3}$B$_{1})$ & 130.48 & 133.05  & 130.56    & 130.39    & 128.93                    & 133.26 \\
SiH$_{3}$               & 222.07 & 220.05  & 222.39    & 222.21    & 220.51                    & 228.08 \\
SiH$_{4}$               & 315.08 & 311.89  & 315.80    & 315.81    & 314.27                    & 324.59 \\
PH$_{2}$                & 148.62 & 146.37  & 147.60    & 146.98    & 144.95                    & 153.97 \\
PH$_{3}$                & 233.35 & 229.18  & 232.06    & 231.69    & 230.24                    & 241.47 \\
SH$_{2}$                & 178.66 & 174.18  & 177.29    & 177.66    & 178.55                    & 183.30  \\
ClH                     & 105.39 & 101.63  & 104.43    & 104.90    & 106.53                    & 107.20 \\
HCCH                    & 406.75 & 399.05  & 403.84    & 405.17    & 409.58                    & 402.76 \\
H$_{2}$CCH$_{2}$        & 561.56 & 554.55  & 559.03    & 559.63    & 561.38                    & 561.34 \\
H$_{3}$CCH$_{3}$        & 708.35 & 701.94  & 706.59    & 706.98    & 707.15                    & 710.20  \\
CN                      & 178.63 & 172.93  & 174.03    & 172.54    & 168.84                    & 180.06 \\
HCN                     & 315.27 & 305.21  & 310.17    & 310.79    & 319.26                    & 311.52 \\
CO                      & 262.93 & 254.60  & 258.48    & 258.64    & 269.29                    & 258.88 \\
HCO                     & 283.19 & 277.00  & 278.81    & 278.50    & 285.79                    & 278.28 \\
H$_{2}$CO               & 375.40 & 367.85  & 370.90    & 370.96    & 379.19                    & 373.21 \\
H$_{3}$COH              & 510.48 & 505.00  & 507.31    & 507.41    & 513.32                    & 511.83 \\
N$_{2}$                 & 229.78 & 218.09  & 223.07    & 222.86    & 234.80                    & 227.44 \\
H$_{2}$NNH$_{2}$        & 433.52 & 428.92  & 428.93    & 427.80    & 432.46                    & 436.70  \\
NO                      & 156.24 & 151.17  & 151.16    & 149.94    & 156.94                    & 152.19 \\
O$_{2}$                 & 126.21 & 119.73  & 120.29    & 119.49    & 128.55                    & 120.54 \\
HOOH                    & 267.02 & 259.77  & 260.59    & 260.11    & 272.51                    & 268.65 \\
F$_{2}$                 & 38.66  & 31.64   & 32.30     & 31.17     & 42.18                     & 38.75  \\
CO$_{2}$                & 400.49 & 390.46  & 393.18    & 393.30    & 409.33                    & 388.59 \\
Si$_{2}$                & 71.64  & 67.21   & 69.78     & 70.72     & 70.56                     & 73.41  \\
P$_{2}$                 & 112.91 & 107.27  & 111.26    & 112.87    & 113.59                    & 115.95 \\
S$_{2}$                 & 104.29 & 100.90  & 102.71    & 103.56    & 103.67                    & 103.11 \\
Cl$_{2}$                & 58.97  & 54.19   & 57.31     & 57.94     & 60.43                     & 59.07  \\
SiO                     & 192.77 & 185.82  & 189.17    & 189.93    & 200.09                    & 192.36 \\
SC                      & 172.35 & 163.07  & 164.01    & 170.43    & 175.16                    & 170.98 \\
SO                      & 127.74 & 122.46  & 123.89    & 123.77    & 129.29                    & 125.80  \\
ClO                     & 62.96  & 60.81   & 59.82     & 58.55     & 59.69                     & 64.53  \\
ClF                     & 62.43  & 57.94   & 58.98     & 58.50     & 65.20                     & 62.57  \\
Si$_{2}$H$_{6}$         & 521.08 & 517.07  & 522.83    & 522.88    & 519.17                    & 535.47 \\
CH$_{3}$Cl              & 393.93 & 388.23  & 392.44    & 393.14    & 394.57                    & 394.52 \\
CH$_{3}$SH              & 470.26 & 463.90  & 468.46    & 469.10    & 469.94                    & 473.49 \\
HOCl                    & 164.83 & 158.46  & 160.70    & 160.75    & 168.50                    & 165.79 \\
SO$_{2}$                & 260.22 & 244.46  & 250.91    & 251.27    & 270.72                    & 259.77 \\
\hline\\[-0.1cm]
MAE                     & 3.19   & 5.49    & 4.31      & 4.31      & 5.37                      &        \\
ME                      & -1.18  & -6.32   & -3.98     & -3.97     & -0.24                     &        \\
RMSD                    & 4.98   & 7.41    & 5.13      & 5.18      & 6.75                      &        \\
Min error               & -14.39 & -18.40  & -12.64    & -12.59    & -16.30                    &        \\
Max error               & 11.90  & 1.87    & 4.59      & 4.71      & 20.74                     &        \\
\hline\hline
\end{tabular}
\end{table*}
\endgroup

\begingroup
\squeezetable
\begin{table*}
\renewcommand{\arraystretch}{0.8}
\setlength{\tabcolsep}{0.07cm}
\footnotesize
\caption{Forward (F) and reverse (R) reaction barrier heights (in kcal/mol) of the DBH24/08 set calculated by DS1DH (with the PBE exchange-correlation functional~\cite{PerBurErn-PRL-96}), RSH+MP2, RSDH with approximations 3 and 4 of Sec.~\ref{sec:approx} (with the short-range PBE exchange-correlation functional of Ref.~\onlinecite{GolWerStoLeiGorSav-CP-06}), and MP2. The calculations were carried out using the
aug-cc-pVQZ basis set at QCISD/MG3 geometries and with parameters $(\mu,\l)$ optimized on the AE6+BH6 combined set. The reference values are taken from Ref.~\onlinecite{ZheZhaTru-JCTC-09}.}
\label{tab:DBH24}
\begin{tabular}{lcccccc}
\hline\hline\\[-0.1cm]
Reaction                                     & DS1DH          & RSH+MP2     & RSDH approx3    & RSDH approx4    & MP2         & Reference   \\
\phantom{xxxxxxxxxxxxxxxxx} $(\mu$,$\l)=$               & (0,0.70)       & (0.58,0)    & (0.46,0.58) & (0.62,0.60) &             &             \\
\hline\\[-0.1cm]
                             &   F/R    &  F/R  &   F/R    & F/R   & F/R & F/R \\
\\
Heavy-atom transfer          &                &             &             &             &             &             \\
H + N$_{2}$O $\rightarrow$ OH + N$_{2}$       & 21.64/75.80    & 19.34/77.14 & 22.76/80.39 & 25.01/82.80 & 35.94/89.26 & 17.13/82.47 \\
H + ClH $\rightarrow$ HCl + H                 & 18.51/18.51    & 19.77/19.77 & 20.23/20.23 & 20.99/20.99 & 22.79/22.79 & 18.00/18.00 \\
CH$_{3}$ + FCl $\rightarrow$ CH$_{3}$F + Cl   &  7.54/60.77    & 8.21/63.59  & 9.79/64.81  & 11.25/66.64 & 19.74/74.29 & 6.75/60.00  \\
\\
Nucleophilic substitution    &                &             &             &             &             &             \\
Cl$^{-}$$\cdots$CH$_{3}$Cl $\rightarrow$ ClCH$_{3}$$\cdots$Cl$^-$ & 12.45/12.45    & 15.40/15.40 & 14.36/14.36 & 9.90/9.90 & 14.64/14.64 & 13.41/13.41 \\
F$^{-}$$\cdots$CH$_{3}$Cl $\rightarrow$ FCH$_{3}$$\cdots$Cl$^{-}$ & 2.83/27.93     & 4.72/31.46  & 3.99/30.52  & 4.27/30.67  & 4.59/28.88  & 3.44/29.42  \\
OH$^{-}$ + CH$_{3}$F $\rightarrow$ HOCH$_{3}$ + F$^{-}$  & -3.11/16.76    & -1.59/21.56 & -1.92/19.23 & -1.53/19.56 & -1.75/17.86 & -2.44/17.66 \\
\\
Unimolecular and association &                &             &             &             &             &             \\
H + N$_{2}$ $\rightarrow$ HN$_{2}$            & 16.36/10.27     & 14.03/13.09 & 17.00/11.57 & 18.75/11.45 & 27.60/8.06 & 14.36/10.61 \\
H + C$_{2}$H$_{4}$ $\rightarrow$ CH$_{3}$CH$_{2}$ & 4.15/44.13  & 2.70/45.76  & 4.34/45.49  & 5.40/45.89  & 9.32/46.54  & 1.72/41.75  \\
HCN $\rightarrow$ HNC                         & 49.13/33.01    & 48.52/34.81 & 49.07/33.59 & 50.05/33.95 & 34.46/52.09 & 48.07/32.82 \\
\\
Hydrogen transfer            &                &             &             &             &             &             \\
OH + CH$_{4}$ $\rightarrow$ CH$_{3}$ + H$_{2}$O  & 4.54/19.33     & 6.03/19.75  & 6.53/20.38  & 7.33/21.35  & 7.66/25.01  & 6.70/19.60  \\
H + OH $\rightarrow$ O + H$_{2}$              & 12.22/11.02    & 13.44/10.00 & 13.49/12.64 & 14.47/13.94 & 17.56/15.58 & 10.70/13.10 \\
H + H$_{2}$S $\rightarrow$ H$_{2}$+ HS        & 4.04/15.09     & 4.73/15.35  & 5.00/15.81  & 5.46/15.96  & 6.42/16.36  & 3.60/17.30  \\
\hline\\[-0.1cm]
MAE                                           & 1.52           & 2.01        & 1.85        & 2.65        & 6.17        &             \\
ME                                            & -0.09          & 1.06        & 1.50        & 1.95        & 4.70        &             \\
RMSD                                          & 2.09           & 2.36        & 2.30        & 3.26        & 8.56        &             \\
Min error                                     & -6.67          & -5.33       & -2.08       & -3.51       & -13.61       &             \\
Max error                                     & 4.51           & 4.01        & 5.63        & 7.88        & 19.61       &            \\
\hline\hline
\end{tabular}
\end{table*}
\endgroup

Finally, the RSDH scheme was tested on the S22 set of 22 weakly interacting molecular complexes~\cite{JurSpoCerHob-PCCP-06}. These calculations were performed with the aug-cc-pVDZ and aug-cc-pVTZ basis sets and the counterpoise correction, using the geometries from Ref.~\onlinecite{JurSpoCerHob-PCCP-06} and the complete-basis-set (CBS)-extrapolated CCSD(T) reference interaction energies from Ref.~\onlinecite{TakHohMalMarShe-JCP-10}. The local MP2 approach~\cite{SchHetWer-JCP-99} is used on the largest systems in the S22 set.

Core electrons are kept frozen in all our MP2 calculations. Spin-restricted calculations are performed for all the closed-shell systems, and spin-unrestricted calculations for all the open-shell systems.

As statistical measures of goodness of the different methods, we compute mean absolute errors (MAEs), mean errors (MEs), root mean square deviations (RMSDs), mean absolute percentage errors (MA\%E), and maximal and minimal errors.

\section{Results and discussion}
\label{sec:results}

\subsection{Optimization of the parameters on the AE6 and BH6 sets}

We start by applying the RSDH scheme on the small AE6 and BH6 sets, and determining optimal values for the parameters $\mu$ and $\l$. Figure~\ref{AE6_BH6} shows the MAEs for these two sets obtained with approximations 1 to 5 of Sec.~\ref{sec:approx} as a function of $\l$ for $\mu=0.5$ and $\mu=0.6$. We choose to show plots for only these two values of $\mu$, since they are close to the optimal value of $\mu$ for RSH+MP2~\cite{ShaTouSav-JCP-11,MusReiAngTou-JCP-15} and also for RSDH with all the approximations except approximation 2. This last approximation is anyhow of little value for thermochemistry since it gives large MAEs on the AE6 set for intermediate values of $\lambda$, which must be related to the incorrect linear dependence in $\l$ of this approximation in the limit $\mu\to 0$ or the high-density limit. We thus only discuss next the other four approximations.

Let us start by analyzing the results for the AE6 set. For the approximations 1, 3, 4, and 5, we can always find an intermediate value of $\lambda$ giving a smaller MAE than the two limiting cases $\l=0$ (corresponding to RSH+MP2) and $\l=1$ (corresponding to standard MP2). Among these four approximations, the approximations 1 and 5 are the least effective to reduce the MAE in comparison to RSH+MP2 and MP2, which may be connected to the fact that these two approximations are both incorrect in the low-density limit. The approximations 3 and 4, which become identical in the high- and low-density limits (for systems with non-zero gaps), are the two best approximations, giving minimal MAEs of 2.2 and 2.3 kcal/mol, respectively, at the optimal parameter values $(\mu,\l)=(0.5,0.6)$ and $(0.6,0.65)$, respectively. 

Let us consider now the results for the BH6 set. Each MAE curve displays a marked minimum at an intermediate value of $\l$, at which the corresponding approximation is globally more accurate than both RSH+MP2 and MP2. All the approximations perform rather similarly, giving minimal MAEs of about 1 kcal/mol. In fact, for $\mu=0.5$ and $\mu=0.6$, the approximations 3 and 4 give essentially identical MAEs for all $\l$. The optimal parameter values for these two approximations are $(\mu,\l)=(0.5,0.5)$, i.e. relatively close to the optimal values found for the AE6 set.

For each of our two best approximations 3 and 4, we also determine optimal values of $\mu$ and $\l$ that minimize the total MAE of the combined AE6 + BH6 set, and which could be used for general chemical applications. For the approximation 3, the optimal parameter values are $(\mu,\l)=(0.46,0.58)$, giving a total MAE of 1.68 kcal/mol. For the approximation 4, the optimal parameter values are $(\mu,\l)=(0.62,0.60)$, giving a total MAE of 1.98 kcal/mol. In the following, we further assess the approximations 3 and 4 with these optimal parameters.

\begin{figure*}[t]
\centering
   \includegraphics[scale=0.40,angle=0]{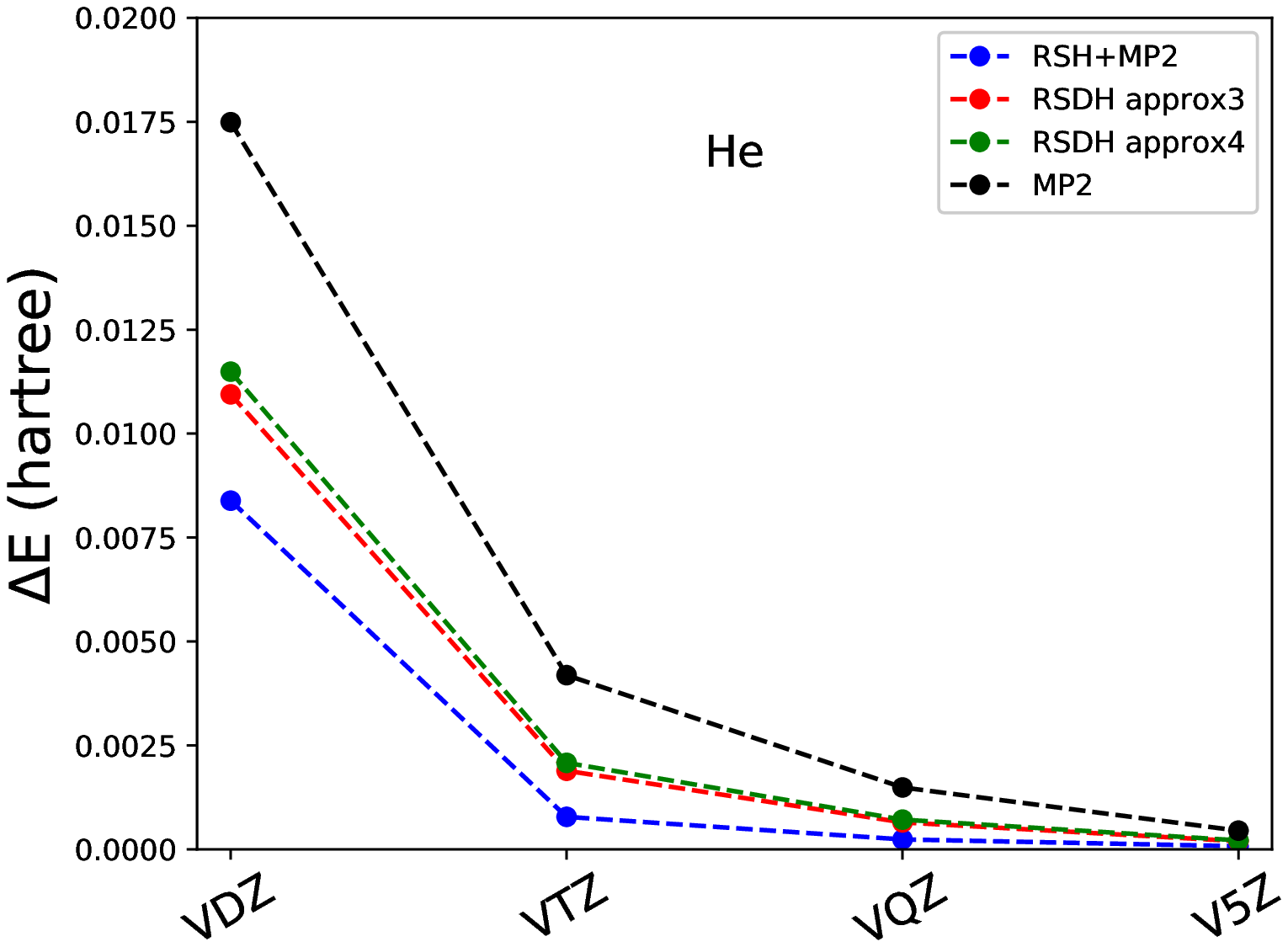}
   \includegraphics[scale=0.40,angle=0]{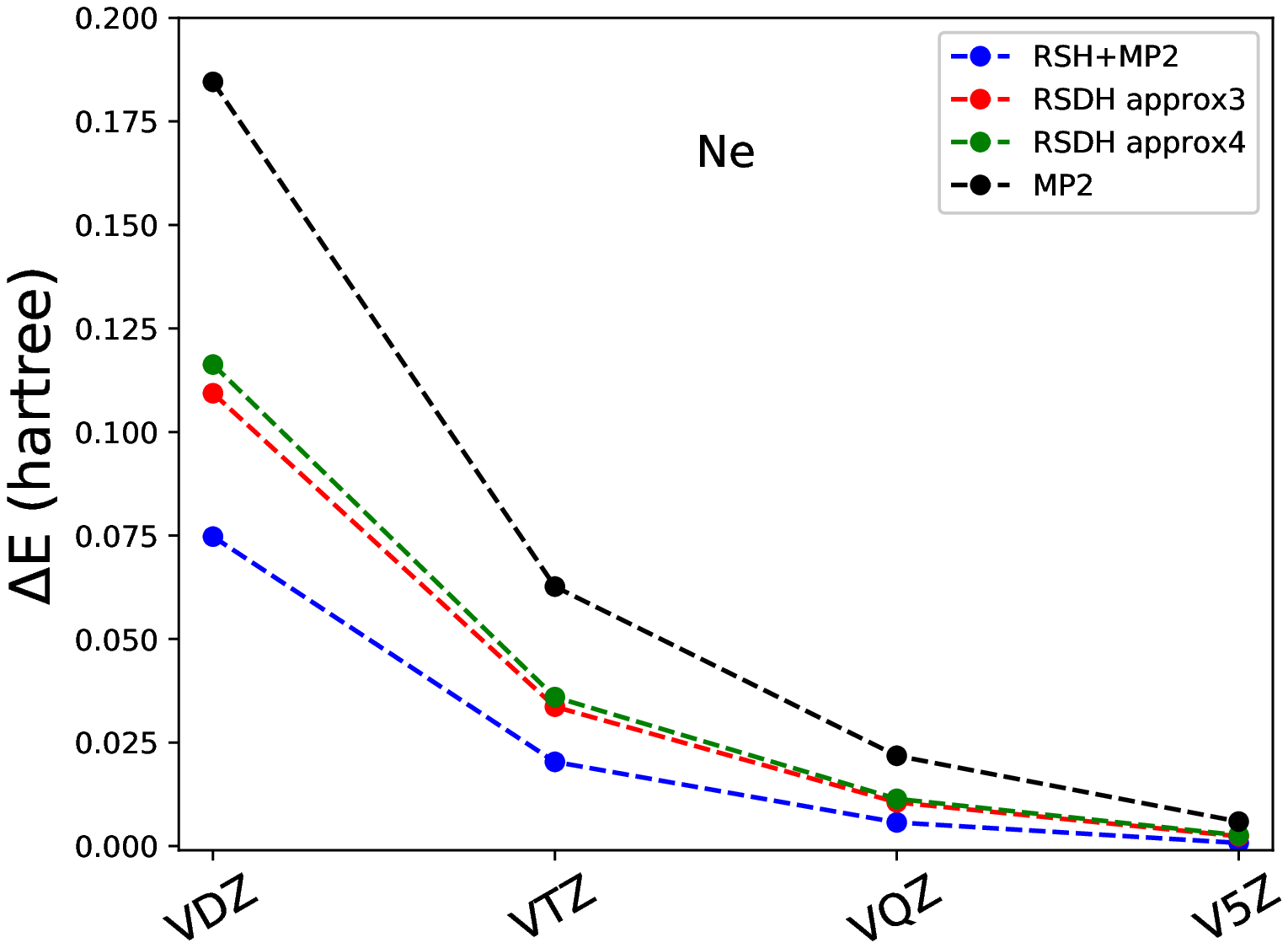} \\
   \includegraphics[scale=0.40,angle=0]{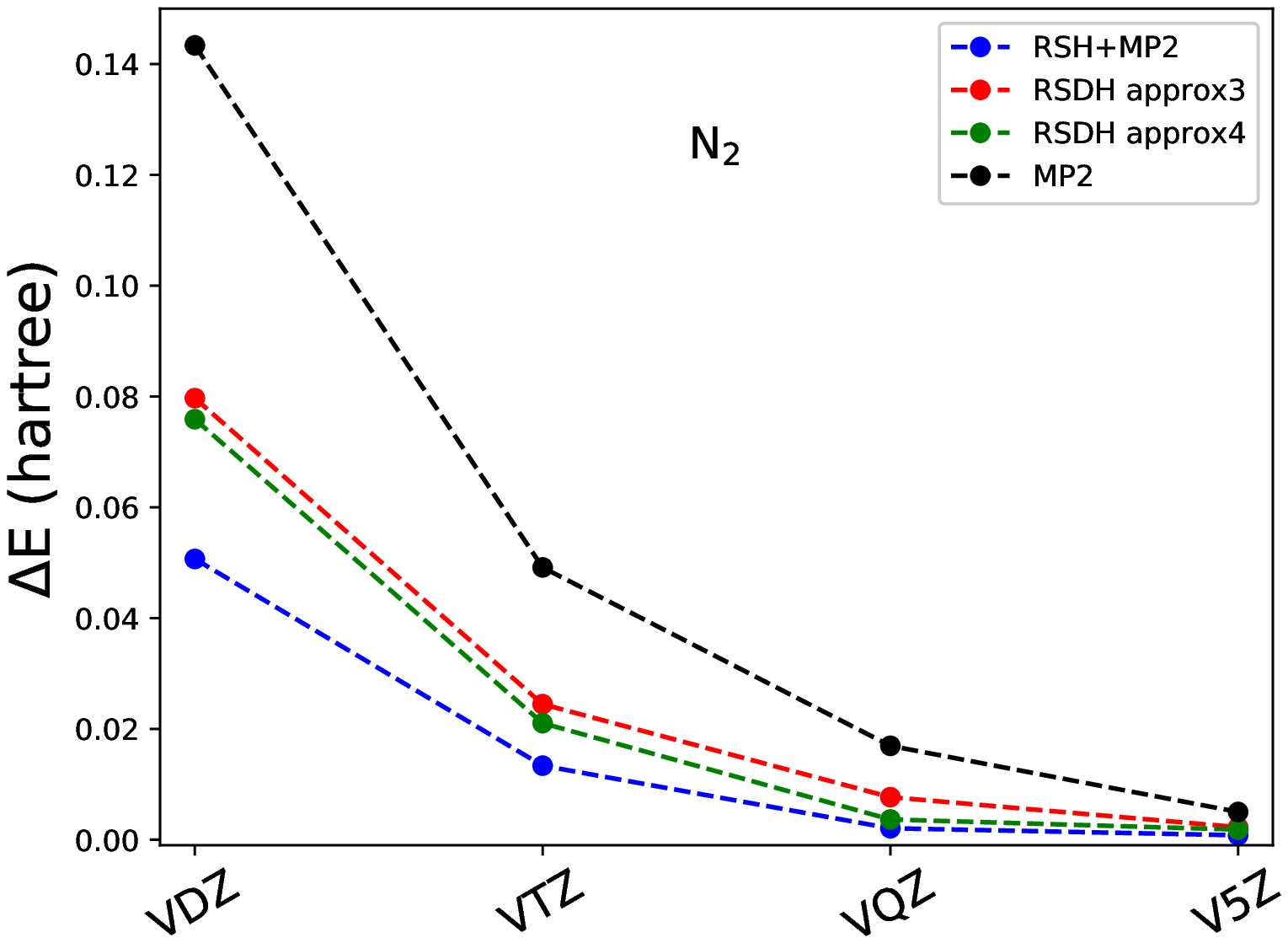}
   \includegraphics[scale=0.40,angle=0]{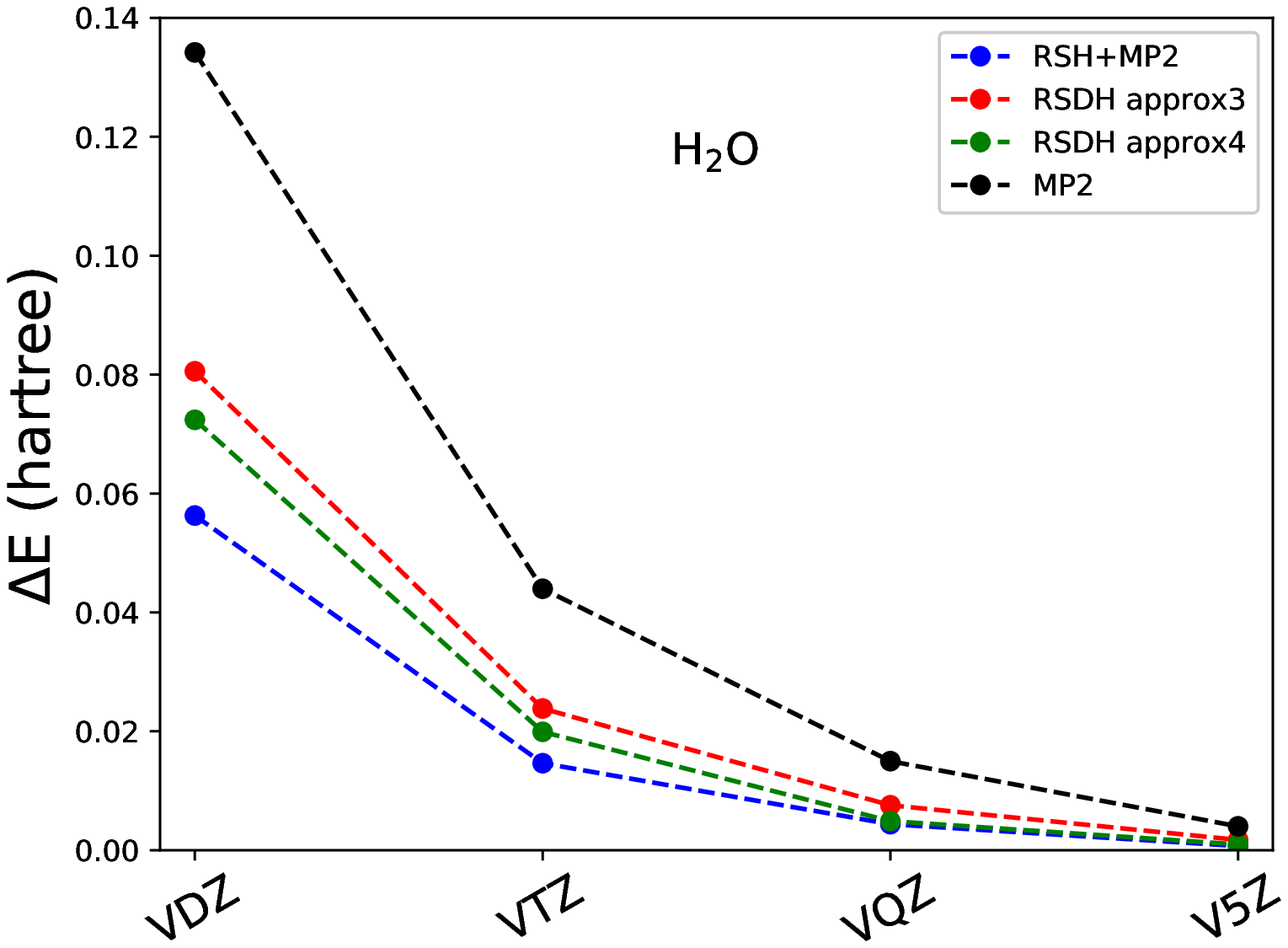}
        \caption{Convergence of the total energy with respect to the basis set for He, Ne, N$_2$, and H$_2$O, as measured by the basis error with respect to the V6Z basis set, $\Delta E = E_{\text{V$X$Z}} - E_\text{V6Z}$, where V$X$Z stands for cc-pV$X$Z with $X=2 \; (\text{D}),3 \; (\text{T}), 4 \; (\text{Q}),5$, calculated with RSH+MP2, RSDH with approximations 3 and 4 of Sec.~\ref{sec:approx} (with the short-range PBE exchange-correlation functional of Ref.~\onlinecite{GolWerStoLeiGorSav-CP-06}), and MP2. The parameters $(\mu,\l)$ used are the ones optimized on the AE6+BH6 combined set.}
        \label{convergence}
\end{figure*}

\subsection{Assessment on the AE49 and DBH24/08 sets of atomization energies and reaction barrier heights}

We assess now the RSDH scheme with the approximations 3 and 4, evaluated with the previously determined optimal parameters $(\mu,\l$), on the larger AE49 and DBH24/08 sets of atomization energies and reaction barrier heights. The results are reported in Tables~\ref{tab:AE49} and~\ref{tab:DBH24}, and compared with other methods corresponding to limiting cases of the RSDH scheme: DS1DH~\cite{ShaTouSav-JCP-11} (with the PBE exchange-correlation functional~\cite{PerBurErn-PRL-96}) corresponding to the $\mu=0$ limit of the RSDH scheme with approximation 4, RSH+MP2~\cite{AngGerSavTou-PRA-05} corresponding to the $\l=0$ limit of the RSDH scheme, and standard MP2 corresponding to the $\mu\to\infty$ or $\l=1$ limit of the RSDH scheme.

On the AE49 set, the two RSDH approximations (3 and 4) give very similar results. With a MAE of 4.3 kcal/mol and a RMSD of about 5.1 kcal/mol, they provide an overall improvement over both RSH+MP2 and standard MP2 which give MAEs larger by about 1 kcal/mol and RMSDs larger by about 2 kcal/mol. It turns out that the DS1DH approximation gives a smaller MAE of 3.2 kcal/mol than the two RSDH approximations, but a similar RMSD of 5.0 kcal/mol. On the DBH24/08 set, the two RSDH approximations give less similar but still comparable results with MAEs of 1.9 and 2.7 kcal/mol for approximations 3 and 4, respectively. This is a big improvement over standard MP2 which gives a MAE of 6.2 kcal/mol, but similar to the accuracy of RSH+MP2 which gives a MAE of 2.0 kcal/mol. Again, the smallest MAE of 1.5 kcal/mol is obtained with the the DS1DH approximation. 

The fact that the DS1DH approximation appears to be globally more accurate that the RSDH approximations on these larger sets but not on the small AE6 and BH6 sets points to a limited representativeness of the latter small sets, and suggests that there may be room for improvement by optimizing the parameters on larger sets.

\subsection{Assessment of the basis convergence}
\label{sec:conv}

We study now the basis convergence of the RSDH scheme. Figure~\ref{convergence} shows the convergence of the total energy of He, Ne, N$_2$, and H$_2$O with respect to the cardinal number $X$ for a series of Dunning basis sets cc-pV$X$Z ($X=2,3,4,5$), calculated with MP2, RSH+MP2, and RSDH with approximations 3 and 4 (with the parameters $(\mu,\l)$ optimized on the AE6+BH6 combined set). 

The results for MP2 and RSH+MP2 are in agreement with what is already known. MP2 has a slow basis convergence, with the error on the total energy decreasing as a third-power law, $\Delta E_{\MP} \sim A \; X^{-3}$~\cite{HelKloKocNog-JCP-97,HalHelJorKloKocOlsWil-CPL-98}, due to the difficulty of describing the short-range part of the correlation hole near the electron-electron cusp. RSH+MP2 has a fast basis convergence, with the error decreasing as an exponential law, $\Delta E_{\text{RSH+MP2}} \sim B \; e^{-\beta X}$~\cite{FraMusLupTou-JCP-15}, since it involves only the long-range MP2 correlation energy.

Unsurprisingly, the RSDH scheme displays a basis convergence which is intermediate between that of MP2 and RSH+MP2. What should be remarked is that, for a given basis, the RSDH basis error is closer to the RSH+MP2 basis error than to the MP2 basis error. The basis dependence of RSDH is thus only moderately affected by the presence of short-range MP2 correlation. This can be understood by the fact that RSDH contains only a modest fraction $\l^2 \approx 0.35$ of the pure short-range MP2 correlation energy $E_{\c,\MP}^{\sr,\mu}$ [see Eq.~(\ref{Emul2decomp})], which should have a third-power-law convergence, while the pure long-range correlation energy $E_{\c,\MP}^{\lr,\mu}$ and the mixed long-range/short-range correlation energy $E_{\c,\MP}^{\lr-\sr,\mu}$ both should have an exponential-law convergence. We thus expect the RSDH error to decrease as $\Delta E_{\text{RSDH}} \sim \l^2 A \; X^{-3} + B \; e^{-\beta X}$, with constants $A$, $B$, $\beta$ \textit{a priori} different from the ones introduced for MP2 and RSH+MP2. The results of Figure~\ref{convergence} are in fact in agreement with such a basis dependence with similar constants $A$, $B$, $\beta$ for MP2, RSH+MP2, and RSDH.

\begingroup
\squeezetable
\begin{table*}
\renewcommand{\arraystretch}{1.0}
\setlength{\tabcolsep}{0.07cm}
\footnotesize
\caption{Interaction energies (in kcal/mol) for the complexes of the S22 set calculated by DS1DH (with the PBE exchange-correlation functional~\cite{PerBurErn-PRL-96}), RSH+MP2, RSDH with approximations 3 and 4 of Sec.~\ref{sec:approx} (with the short-range PBE exchange-correlation functional of Ref.~\onlinecite{GolWerStoLeiGorSav-CP-06}), and MP2. The parameters $(\mu,\l)$ used are the ones optimized on the AE6+BH6 combined set, except for the RSH+MP2 values which are taken from Ref.~\onlinecite{ZhuTouSavAng-JCP-10} in which $\mu=0.50$ was used. The basis sets used are aVDZ and aVTZ which refer to aug-cc-pVDZ and aug-cc-pVTZ, respectively, and the counterpoise correction is applied. The values in italics were obtained using the local MP2 approach, the ones with an asterisk (*) were obtained in Ref.~\onlinecite{GolLeiManMitWerSto-PCCP-08} with the density-fitting approximation, and the ones with a dagger ($\dagger$) were obtained with the approximation: $E_{\text{aVTZ}}(\text{RSDH approx4}) \approx E_{\text{aVDZ}}(\text{RSDH approx4}) + E_{\text{aVTZ}}(\text{RSDH approx3}) - E_{\text{aVDZ}}(\text{RSDH approx3})$. The geometries of the complexes are taken from Ref.~\onlinecite{JurSpoCerHob-PCCP-06} and the reference interaction energies are taken as the CCSD(T)/CBS estimates of Ref.~\onlinecite{TakHohMalMarShe-JCP-10}. The MP2 values are also taken from Ref.~\onlinecite{TakHohMalMarShe-JCP-10}.}
\label{tab:S22}
\begin{tabular}{lcccccccccc}
\hline\hline\\[-0.1cm]
Complex         & \phantom{x} DS1DH \phantom{x}   & \multicolumn{2}{c}{RSH+MP2}           &  \multicolumn{2}{c}{RSDH approx3}             & \multicolumn{2}{c}{RSDH approx4}     & \multicolumn{2}{c}{MP2} & Reference    \\
\phantom{xxxxxxxxxxxxx} $(\mu $,$\l)=$ & (0,0.70) & \multicolumn{2}{c}{(0.50,0)}          &  \multicolumn{2}{c}{(0.46,0.58)}              & \multicolumn{2}{c}{(0.62,0.60)}      &            &            &         \\
                           & aVDZ               & aVDZ          & aVTZ                  & aVDZ               & aVTZ             & aVDZ                  & aVTZ                 & aVDZ       & aVTZ       &         \\
\hline\\[-0.1cm]
\multicolumn{5}{l}{Hydrogen-bonded complexes}\\[0.1cm]
Ammonia dimer                & -2.70              & -3.13         & -3.25                 & -3.00              &   -3.18          & -2.94                 &   -3.16              & -2.68      & -2.99      & -3.17    \\
Water dimer                  & -4.63              & -5.34         & -5.45                 & -5.03              &   -5.19          & -4.93                 &   -5.12              & -4.36      & -4.69      & -5.02    \\
Formic acid dimer            & -17.28             & -21.20        & -21.57                & -19.31             &   -20.14         & -18.86                &   -19.80             & -15.99     & -17.55     & -18.80   \\
Formamide dimer              & -14.63             & -17.44        & -17.64                & -16.30             &   -16.81         & -15.98                &   -16.60             & -13.95     & -15.03     & -16.12   \\
Uracile dimer $C_{2h}$       & \textit{-18.86}    & -22.62        & \textit{-22.82}*      & \textit{-20.52}    &  \textit{-21.77} & \textit{-20.53}       &   \textit{-21.78}$^\dagger$  & -18.41     & -19.60     & -20.69   \\
2-pyridoxine/2-aminopyridine & \textit{-18.65}    & -18.86        & \textit{-18.60}*      & \textit{-17.43}    &  \textit{-17.93} & \textit{-17.04}       &   \textit{-17.55}$^\dagger$  & -15.56     & -16.64     & -17.00   \\
Adenine/thymine WC           & \textit{-17.52}    & -18.26        & \textit{-18.12}*      & \textit{-16.47}    &  \textit{-17.28} & \textit{-16.23}       &   \textit{-17.04}$^\dagger$  & -14.71     & -15.80     & -16.74   \\
\hline\\[-0.3cm]
MAE                          & 1.16               &  1.34         & 1.42                  & 0.26               & 0.68             & 0.23                  &       0.51           & 1.70       & 0.75       &         \\
ME                           & 0.46               & -1.33         & -1.42                 & -0.07              & -0.68            & 0.22                  &      -0.50           & 1.70       & 0.75       &         \\
RMSD                         & 1.28               &  1.56         & 1.66                  & 0.29               & 0.81             & 0.29                  &       0.64           & 1.88       & 0.85       &         \\
MA\%E                        & 9.00               &  8.36         & 9.03                  & 2.04               & 4.14             & 2.35                  &       3.01           & 12.63      & 5.62       &         \\
\hline \\[-0.2cm]
\multicolumn{5}{l}{Complexes with predominant dispersion contribution}\\[0.1cm]
Methane dimer                & -0.25              & -0.46          & -0.48                & -0.42              & -0.47            & -0.42                 &   -0.47              & -0.39       & -0.46      & -0.53    \\
Ethene dimer                 & -0.84              & -1.45          & -1.55                & -1.38              & -1.68            & -1.33                 &   -1.55              & -1.18       & -1.46      & -1.50    \\
Benzene/methane              & -0.87              & -1.62          & -1.71                & -1.56              & -1.70            & -1.56                 &   -1.63              & -1.47       & -1.71      & -1.45    \\
Benzene dimer $C_{2h}$       & -7.21              & -4.08          &\textit{-4.24}*       & -3.52              & -3.78            & -4.14                 &   -4.40$^\dagger$            & -4.25       & -4.70      & -2.62    \\
Pyrazine dimer               & -8.97              & -5.97          &\textit{-6.04}*       & -6.50              & -6.21            & -6.02                 &   -5.73$^\dagger$            & -6.00       & -6.55      & -4.20    \\
Uracil dimer $C_{2}$         & -13.31             & -11.76         &\textit{-11.95}*      & -12.70             & -11.42           & -10.77                &   -9.49$^\dagger$            & -9.80       & -10.63     & -9.74    \\
Indole/benzene               & \textit{-17.26}    & -6.95          &\textit{-6.96}*       & \textit{-8.83}     & \textit{-6.97}   & \textit{-9.25}        &   \textit{-7.39}$^\dagger$   & -7.13       & -7.74      & -4.59    \\
Adenine/thymine stack        & \textit{-20.84}    & -15.11         &\textit{-14.71}*      & \textit{-14.28}    & \textit{-14.56}  & \textit{-14.25}       &   \textit{-14.53}$^\dagger$  & -13.24      & -14.26     & -11.66   \\
\hline \\[-0.3cm]
MAE                          & 4.54               & 1.42           & 1.43                 & 1.67               & 1.33             & 1.50                  & 1.19                 & 1.01        & 1.43       &         \\
ME                           & -4.16              & -1.39          & -1.42                & -1.61              & -1.31            & -1.43                 & -1.11                & -0.90       & -1.40      &         \\
RMSD                         & 6.14               & 1.83           & 1.80                 & 2.23               & 1.67             & 2.10                  & 1.65                 & 1.37        & 1.85       &         \\
MA\%E                        & 102.09             & 28.48          & 29.60                & 34.02              & 28.31            & 34.42                 & 27.45                & 27.96       & 33.65      &         \\
\hline \\[-0.2cm]
\multicolumn{5}{l}{Mixed complexes}\\[0.1cm]
Ethene/ethyne                & -1.28             & -1.62          & -1.68                & -1.57               & -1.68             & -1.43                 &    -1.67              & -1.39      & -1.58      & -1.51    \\
Benzene/water                & -2.66             & -3.49          & -3.68                & -3.33               & -3.55             & -3.29                 &    -3.53              & -2.98      & -3.35      & -3.29    \\
Benzene/ammonia              & -1.70             & -2.49          & -2.63                & -2.39               & -2.58             & -2.38                 &    -2.59              & -2.21      & -2.52      & -2.32    \\
Benzene/hydrogen cyanide     & -3.86             & -5.31          & -5.38                & -4.93               & -5.26             & -4.89                 &    -5.26              & -4.37      & -4.92      & -4.55    \\
Benzene dimer  C$_{2v}$      & -4.57             & -3.33          & \textit{-3.49}*      & -3.26               & -3.47             & -3.26                 &    -3.47$^\dagger$            & -3.09      & -3.46      & -2.71    \\
Indole/benzene T-shaped      &\textit{-11.71}    & -6.55          & \textit{-6.85}*      &\textit{-7.49}       & \textit{-6.50}    &\textit{-7.91}         &    \textit{-6.92}$^\dagger$   & -6.10      & -6.71      & -5.62    \\
Phenol dimer                 &\textit{-8.05}     & -8.05          & \textit{-8.09}*      & \textit{-6.89}      & \textit{-7.57}    &\textit{-7.15}         &    \textit{-7.83}$^\dagger$   & -6.79      & -7.36      & -7.09    \\
\hline\\[-0.3cm]
MAE                          & 1.58              & 0.54           & 0.67                 & 0.45                & 0.50              & 0.48                  &     0.60              & 0.27       & 0.40       &         \\
ME                           & -0.97             & -0.54          & -0.67                & -0.40               & -0.50             & -0.46                 &    -0.60              &  0.02      & -0.40      &         \\
RMSD                         & 2.31              & 0.76           & 0.86                 & 0.75                & 0.57              & 0.90                  &     0.71              & 0.50       & 0.69       &         \\
MA\%E                        & 38.01             & 12.34          & 17.40                & 10.43               & 13.78             & 11.03                 &     15.28             & 7.55       & 10.58      &         \\
\hline\\[-0.2cm]
Total MAE                    & 2.52              & 1.11           & 1.18                 & 0.83                &0.86               & 0.77                  &       0.75             & 0.99       & 0.89       &         \\
Total ME                     & -1.67             & -1.09          & -1.18                & -0.73               &-0.85              & -0.60                 &      -0.75             &  0.22      & -0.40      &         \\
Total RMSD                   & 4.03              & 1.45           & 1.49                 & 1.42                &1.15               & 1.37                  &       1.13             & 1.35       & 1.25       &          \\
Total MA\%E                  & 52.08             & 16.95          & 19.06                & 16.34               &16.00              & 16.77                 &       15.80            & 16.59      & 17.36      &          \\
\hline\hline
\end{tabular}
\end{table*}
\endgroup

\subsection{Assessment on the S22 set of intermolecular interactions}

We finally test the RSDH scheme on weak intermolecular interactions. Table~\ref{tab:S22} reports the interaction energies for the 22 molecular dimers of the S22 set calculated by RSH+MP2, RSDH (with approximations 3 and 4), and MP2, using the aug-cc-pVDZ and aug-cc-pVTZ basis sets. We also report DS1DH results, but since this method is quite inaccurate for dispersion interactions we only did calculations with the aug-cc-pVDZ basis set for a rough comparison. Again, the basis dependence of RSDH is intermediate between the small basis dependence of RSH+MP2 and the larger basis dependence of standard MP2. The basis convergence study in Section~\ref{sec:conv} suggests that the RSDH results with the aug-cc-pVTZ basis set are not far from the CBS limit.

The two approximations (3 and 4) used in the RSDH scheme give overall similar results, which may be rationalized by the fact low-density regions primarily contribute to these intermolecular interaction energies and the approximations 3 and 4 become identical in the low-density limit. For hydrogen-bonded complexes, RSDH with the aug-cc-pVTZ basis set gives a MA\%E of about 3-4\%, similar to standard MP2 but in clear improvement over RSH+MP2 which tends to give too negative interaction energies. Presumably, this is so because the explicit wave-function treatment of the short-range interaction $\l w_{\ee}^{\sr,\mu}(r_{12})$ makes RSDH accurately describe of the short-range component of the intermolecular interaction, while still correctly describe the long-range component. For complexes with a predominant dispersion contribution, RSDH with the aug-cc-pVTZ basis set gives too negative interaction energies by about 30 \%, similar to both MP2 and RSH+MP2. Notably, DS1DH gives much too negative interaction energies for the largest and most polarizable systems, leading to a MA\%E of more than 100 \% with aug-cc-pVDZ basis set. This can be explained by the fact that the reduced amount of HF exchange at long range in DS1DH leads to smaller HOMO-LUMO gaps in these systems in comparison with RSH+MP2 and RSDH, causing a overlarge MP2 contribution. For mixed complexes, RSDH with the aug-cc-pVTZ basis set gives a MA\%E of about 14-15 \%, which is a bit worse than MP2 but slightly better than RSH+MP2. Again, DS1DH tends to give significantly too negative interaction energies for the largest dimers.

Overall, for weak intermolecular interactions, RSDH thus provides a big improvement over DS1DH, a small improvement over RSH+MP2, and is quite similar to standard MP2.

\section{Conclusion}
\label{sec:conclusion}

We have studied a wave-function/DFT hybrid approach based on a CAM-like decomposition of the electron-electron interaction in which a correlated wave-function calculation associated with the two-parameter interaction $w_\ee^{\lr,\mu} (r_{12}) + \l w_\ee^{\sr,\mu} (r_{12})$ is combined with a complement short-range density functional. Specifically, we considered the case of MP2 perturbation theory for the wave-function part and obtained a scheme that we named RSDH. This RSDH scheme is a generalization of the usual one-parameter DHs (corresponding to the special case $\mu=0$) and the range-separated MP2/DFT hybrid known as RSH+MP2 (corresponding to the special case $\l=0$). It allows one to have both 100\% HF exchange and MP2 correlation at long interelectronic distances and fractions of HF exchange and MP2 correlation at short interelectronic distances. We have also proposed a number of approximations for the complement short-range exchange-correlation density functional, based on the limits $\mu=0$ and $\mu\to\infty$, and showed their relevance on the uniform-electron gas with the corresponding electron-electron interaction, in particular in the high- and low-density limits.

The RSDH scheme with complement short-range DFAs constructed from a short-range version of the PBE functional has then been applied on small sets of atomization energies (AE6 set) and reaction barrier heights (BH6 set) in order to find optimal values for the parameters $\mu$ and $\l$. It turns out that the optimal values of these parameters for RSDH, $\mu \approx 0.5-0.6$ and $\l \approx 0.6$, are very similar to the usual optimal values found separately for RSH+MP2 and one-parameter DHs. With these values of the parameters, RSDH has a relatively fast convergence with respect to the size of the one-electron basis, which can be explained by the fact that its contains only a modest fraction $\l^2 \approx 0.35$ of pure short-range MP2 correlation. We have tested the RSDH scheme with the two best complement short-range DFAs (referred to as approximations 3 and 4) on large sets of atomization energies (AE49 set), reaction barrier heights (DBH24 set), and weak intermolecular interactions (S22 set). The results show that the RSDH scheme is either globally more accurate or comparable to RSH+MP2 and standard MP2. If we had to recommend a computational method for general chemical applications among the methods tested in this work, it would be RSDH with approximation 3 with parameters $(\mu,\l)=(0.46,0.58)$.

There is much room however for improvement and extension. The parameters $\mu$ and $\l$ could be optimized on larger training sets. More accurate complement short-range DFAs should be constructed. The MP2 correlation term could be replaced by random-phase approximations, which would more accurately describe dispersion interactions~\cite{ZhuTouSavAng-JCP-10,TouZhuSavJanAng-JCP-11}, or by multireference perturbation theory~\cite{FroCimJen-PRA-10}, which would capture static correlation effects. The RSDH scheme could be extended to linear-response theory for calculating excitation energies or molecular properties, e.g. by generalizing the methods of Refs.~\onlinecite{GriNee-JCP-07,RebSavTou-MP-13,HedHeiKneFroJen-JCP-13,RebTou-JCP-16}. 

\section*{Acknowledgements}
We thank Bastien Mussard for help with the MOLPRO software. We also thank Labex MiChem for providing PhD financial support for C. Kalai.

\appendix*
\begin{widetext}
\section{Uniform coordinate scaling relation and Coulomb/high-density and short-range/low-density limits of $\bar{E}_\c^{\sr,\mu,\l}[n]$}
\label{appendix}

\subsection{Scaling relation for $\bar{E}_\c^{\sr,\mu,\l}[n]$}

Here, we generalize the uniform coordinate scaling relation, known for the KS correlation functional $E_\c[n]$~\cite{LevPer-PRA-85,Lev-PRA-91,Lev-INC-95} and for the complement short-range correlation functional $\bar{E}_\c^{\sr,\mu}[n]$~\cite{Tou-THESIS-05,TouGorSav-IJQC-06}, to the $\l$-dependent complement short-range correlation functional $\bar{E}_\c^{\sr,\mu,\l}[n]$. We first define the universal density functional, for arbitrary parameters $\mu \geq 0$, $\l \geq 0$, and $\xi \geq 0$,
\begin{equation}
F^{\mu,\lambda,\xi}[n] = \min_{\Psi\to n} \bra{\Psi} \hat{T} + \xi \hat{W}_\ee^{\lr,\mu} + \l \hat{W}_\ee^{\sr,\mu} \ket{\Psi},
\label{Fmlx}
\end{equation}
which is a simple generalization of the universal functional $F^{\mu,\lambda}[n]$ in Eq.~(\ref{LL-decomposition2}) such that $F^{\mu,\lambda,\xi=1}[n]=F^{\mu,\lambda}[n]$. The minimizing wave function in Eq.~(\ref{Fmlx}) will be denoted by $\Psi^{\mu,\lambda,\xi}[n]$. Let us now consider the scaled wave function $\Psi_\gamma^{\mu/\gamma,\lambda/\gamma,\xi/\gamma}[n]$ defined by, for $N$ electrons,
\begin{eqnarray}
\Psi_\gamma^{\mu/\gamma,\lambda/\gamma,\xi/\gamma}[n](\b{r}_1,...,\b{r}_N)  = 
\gamma^{3N/2} \; \Psi^{\mu/\gamma,\lambda/\gamma,\xi/\gamma}[n](\gamma \b{r}_1,...,\gamma\b{r}_N),
\end{eqnarray}
where $\gamma>0$ is a scaling factor. The wave function $\Psi_\gamma^{\mu/\gamma,\lambda/\gamma,\xi/\gamma}[n]$ yields the scaled density $n_\gamma(\b{r}) = \gamma^3 n (\gamma \b{r})$ and minimizes $\bra{\Psi} \hat{T} + \xi \hat{W}_\ee^{\lr,\mu} + \l \hat{W}_\ee^{\sr,\mu} \ket{\Psi}$ since
\begin{eqnarray}
\bra{\Psi_\gamma^{\mu/\gamma,\lambda/\gamma,\xi/\gamma}[n]} \hat{T} + \xi \hat{W}_\ee^{\lr,\mu} + \l \hat{W}_\ee^{\sr,\mu} \ket{\Psi_\gamma^{\mu/\gamma,\lambda/\gamma,\xi/\gamma}[n]} = \phantom{xxxxxxxxxxxxxxxxxxx}
\nonumber\\
\gamma^2 \bra{\Psi^{\mu/\gamma,\lambda/\gamma,\xi/\gamma}[n]} \hat{T} + (\xi/\gamma) \hat{W}_\ee^{\lr,\mu/\gamma} + (\l/\gamma) \hat{W}_\ee^{\sr,\mu/\gamma} \ket{\Psi^{\mu/\gamma,\lambda/\gamma,\xi/\gamma}[n]},
\end{eqnarray}
where the right-hand side is minimal by definition of $\Psi^{\mu/\gamma,\lambda/\gamma,\xi/\gamma}[n]$. Therefore, we conclude that
\begin{equation}
\Psi^{\mu,\lambda,\xi}[n_\gamma] = \Psi_\gamma^{\mu/\gamma,\lambda/\gamma,\xi/\gamma}[n],
\end{equation}
and
\begin{equation}
F^{\mu,\lambda,\xi}[n_\gamma] = \gamma^2 F^{\mu/\gamma,\lambda/\gamma,\xi/\gamma}[n].
\end{equation}
Consequently, the corresponding correlation functional,
\begin{eqnarray}
E_\c^{\mu,\lambda,\xi}[n]=\bra{\Psi^{\mu,\lambda,\xi}[n]} \hat{T} + \xi \hat{W}_\ee^{\lr,\mu} + \l \hat{W}_\ee^{\sr,\mu} \ket{\Psi^{\mu,\lambda,\xi}[n]} 
- \bra{\Phi[n]} \hat{T} + \xi \hat{W}_\ee^{\lr,\mu} + \l \hat{W}_\ee^{\sr,\mu} \ket{\Phi[n]},
\label{Ecmlx}
\end{eqnarray}
with the KS single-determinant wave function $\Phi[n]=\Psi^{\mu=0,\lambda=0,\xi}[n]$, satisfies the same scaling relation
\begin{equation}
E_\c^{\mu,\lambda,\xi}[n_\gamma] = \gamma^2 E_\c^{\mu/\gamma,\lambda/\gamma,\xi/\gamma}[n].
\label{app:Ecmlxng}
\end{equation}
Similarly, the associated short-range complement correlation functional,
\begin{equation}
\bar{E}_\c^{\sr,\mu,\lambda,\xi}[n] = E_\c^{\xi}[n] - E_\c^{\mu,\lambda,\xi}[n],
\label{Ecsrmlx}
\end{equation}
with $E_\c^{\xi}[n] = E_\c^{\mu\to\infty,\lambda,\xi}[n]$, satisfies the scaling relation
\begin{equation}
\bar{E}_\c^{\sr,\mu,\lambda,\xi}[n_\gamma] = \gamma^2 \bar{E}_\c^{\sr,\mu/\gamma,\lambda/\gamma,\xi/\gamma}[n].
\label{Ecsrmulxingamma}
\end{equation}
Applying this relation for $\xi=1$ gives the scaling relation for $\bar{E}_\c^{\sr,\mu,\l}[n]$,
\begin{equation}
\bar{E}_\c^{\sr,\mu,\lambda}[n_\gamma] = \gamma^2 \bar{E}_\c^{\sr,\mu/\gamma,\lambda/\gamma,1/\gamma}[n],
\label{Ecsrmlscaling}
\end{equation}
from which we see that the high-density limit $\gamma\to\infty$ is related to the Coulomb limit $\mu\to 0$ and the low-density limit $\gamma\to 0$ is related to the short-range limit $\mu\to \infty$ of $\bar{E}_\c^{\sr,\mu,\l}[n]$.

Note that by applying Eq.~(\ref{Ecsrmulxingamma}) with $\lambda=0$ and $\gamma=\xi$ we obtain the short-range complement correlation functional associated with the interaction $\xi w_\ee^{\lr,\mu}$ in terms of the short-range complement correlation functional associated with the interaction $w_\ee^{\lr,\mu/\xi}$, i.e. $\bar{E}_\c^{\sr,\mu,0,\xi}[n] = \xi^2 \bar{E}_\c^{\sr,\mu/\xi,0,1}[n_{1/\xi}] = \xi^2 \bar{E}_\c^{\sr,\mu/\xi}[n_{1/\xi}]$, as already explained in Ref.~\onlinecite{TouGorSav-IJQC-06}. Also, by applying Eq.~(\ref{Ecsrmulxingamma}) with $\xi=1$ and $\gamma=\lambda$ we obtain the short-range complement correlation functional associated with the interaction $w_\ee^{\lr,\mu}  + \lambda w_\ee^{\sr,\mu}$ in terms of the short-range complement correlation functional associated with the interaction $(1/\lambda) w_\ee^{\lr,\mu/\lambda}  + w_\ee^{\sr,\mu/\lambda}$, i.e. $\bar{E}_\c^{\sr,\mu,\lambda}[n] = \bar{E}_\c^{\sr,\mu,\lambda,1}[n] = \lambda^2 \bar{E}_\c^{\sr,\mu/\lambda,1,1/\lambda}[n_{1/\lambda}]$. However, since the functional $\bar{E}_\c^{\sr,\mu,1,\xi}[n]$ is equally unknown as the functional $\bar{E}_\c^{\sr,\mu,\lambda}[n]$, this relation is not useful to obtain an expression for the functional $\bar{E}_\c^{\sr,\mu,\lambda}[n]$.

\subsection{Coulomb limit and high-density limit of $\bar{E}_\c^{\sr,\mu,\l}[n]$}

We first give the limit of $\bar{E}_\c^{\sr,\mu,\l,\xi}[n]$ as $\mu\to 0$. Starting from Eq.~(\ref{Ecsrmlx}) and noting that $E_\c^{\mu=0,\lambda,\xi}[n] = E_\c^{\lambda}[n]$, we obtain:
\begin{eqnarray}
\bar{E}_\c^{\sr,\mu=0,\l,\xi}[n] = E_\c^{\xi}[n] - E_\c^{\lambda}[n]
                                 = \xi^2 E_\c[n_{1/\xi}] - \l^2 E_\c[n_{1/\l}],
\label{app:barEcsrm0lxi}
\end{eqnarray}
where we have used the well-known relation, $E_\c^{\lambda}[n] =  \l^2 E_\c[n_{1/\l}]$~\cite{Lev-PRA-91,Lev-INC-95,ShaTouSav-JCP-11} [a special case of Eq.~(\ref{app:Ecmlxng})]. In particular, for $\xi=1$ we obtain the limit of $\bar{E}_\c^{\sr,\mu,\l}[n]$ as $\mu\to 0$
\begin{eqnarray}
\bar{E}_\c^{\sr,\mu=0,\l}[n] = E_\c[n] - \l^2 E_\c[n_{1/\l}].
\label{app:barEcl}
\end{eqnarray}

We can now derive the high-density limit of $\bar{E}_\c^{\sr,\mu,\l}[n]$ using the scaling relation in Eq.~(\ref{Ecsrmlscaling}) and the limit $\mu\to 0$ in Eq.~(\ref{app:barEcsrm0lxi})
\begin{eqnarray}
\lim_{\gamma\to \infty}\bar{E}_\c^{\sr,\mu,\lambda}[n_\gamma] 
&=& \lim_{\gamma\to \infty} \gamma^2 \bar{E}_\c^{\sr,\mu/\gamma,\lambda/\gamma,1/\gamma}[n]
\nonumber\\
&=& \lim_{\gamma\to \infty} \left( E_\c[n_{\gamma}] - \l^2 E_\c[n_{\gamma/\l}] \right)
\nonumber\\
&=& \left( 1 - \lambda^2 \right) E_\c^{\text{GL2}}[n],
\end{eqnarray}
where we have used $\lim_{\gamma\to \infty} E_\c[n_{\gamma}] = E_\c^{\text{GL2}}[n]$~\cite{GorLev-PRB-93} assuming a KS system with a non-degenerate ground state.

\subsection{Short-range limit and low-density limit of $\bar{E}_\c^{\sr,\mu,\l}[n]$}

We first derive the leading term of the asymptotic expansion of $\bar{E}_\c^{\sr,\mu,\l,\xi}[n]$ as $\mu\to\infty$. Taking the derivative with respect to $\l$ of Eq.~(\ref{Ecmlx}), and using the Hellmann-Feynman theorem which states that the derivative of $\Psi^{\mu,\l,\xi}[n]$ does not contribute, we obtain:
\begin{eqnarray}
\frac{\partial E_\c^{\mu,\l,\xi}[n]}{\partial \l} &=& \bra{\Psi^{\mu,\l,\xi}[n]} \hat{W}_\ee^{\sr,\mu} \ket{\Psi^{\mu,\l,\xi}[n]} 
- \bra{\Phi[n]} \hat{W}_\ee^{\sr,\mu} \ket{\Phi[n]}
\nonumber\\
&=& \frac{1}{2} \iint  n_{2,\c}^{\mu,\l,\xi}[n](\b{r}_1,\b{r}_2) w_\ee^{\sr,\mu}(r_{12}) \d\b{r}_1\d\b{r}_2,
\end{eqnarray}
where $n_{2,\c}^{\mu,\l,\xi}[n](\b{r}_1,\b{r}_2) = \bra{\Psi^{\mu,\l,\xi}[n]} \hat{n}_2(\b{r}_1,\b{r}_2)\ket{\Psi^{\mu,\l,\xi}[n]} - \bra{\Phi[n]} \hat{n}_2(\b{r}_1,\b{r}_2) \ket{\Phi[n]}$ is the correlation part of the pair density associated with the wave function $\Psi^{\mu,\l,\xi}[n]$. Noting from Eq.~(\ref{Ecsrmlx}) that $\partial \bar{E}_\c^{\sr,\mu,\l,\xi}[n]/\partial \l= - \partial E_\c^{\mu,\l,\xi}[n]/\partial \l$, integrating over $\l$, and using $\bar{E}_\c^{\sr,\mu,\l=\xi,\xi}[n]=0$, we arrive at the exact expression:
\begin{eqnarray}
\bar{E}_\c^{\sr,\mu,\l,\xi}[n]&=& \frac{1}{2} \int_{\l}^\xi \d\alpha 
\iint n_{2,\c}^{\mu,\alpha,\xi}[n](\b{r}_1,\b{r}_2) w_\ee^{\sr,\mu}(r_{12}) \d\b{r}_1\d\b{r}_2.
\label{Ecintl}
\end{eqnarray}
Using now the asymptotic expansion of the short-range interaction~\cite{TouColSav-PRA-04},
\begin{eqnarray}
w_\ee^{\sr,\mu}(r_{12}) = \frac{\pi}{\mu^2} \delta(\b{r}_{12}) + {\cal O}\left( \frac{1}{\mu^3}\right),
\label{asym-short}
\end{eqnarray}
we obtain the leading term of the asymptotic expansion of $\bar{E}_\c^{\sr,\mu,\l,\xi}[n]$ as $\mu\to\infty$,
\begin{eqnarray}
\bar{E}_\c^{\sr,\mu,\l,\xi}[n]&=& (\xi-\l) \frac{\pi}{2\mu^2}  \int n_{2,\c}^{\xi}[n](\b{r},\b{r}) \d\b{r} + \; {\cal O}\left( \frac{1}{\mu^3}\right),
\label{Ecmulxmuinfty}
\end{eqnarray}
where $n_{2,\c}^\xi[n](\b{r},\b{r})= n_{2,\c}^{\mu\to\infty,\alpha,\xi}[n](\b{r},\b{r})$ is the correlation part of the on-top pair density associated with the scaled Coulomb interaction $\xi  w_\ee(r_{12})$. For the special case $\xi=1$, we obtain the leading term of the asymptotic expansion of $\bar{E}_\c^{\sr,\mu,\l}[n]$
\begin{eqnarray}
\bar{E}_\c^{\sr,\mu,\l}[n]&=& (1-\l) \frac{\pi}{2\mu^2}  \int n_{2,\c}[n](\b{r},\b{r}) \d\b{r} + \; {\cal O}\left( \frac{1}{\mu^3}\right),
\label{app:Ecmulmuinfty}
\end{eqnarray}
where $n_{2,\c}[n](\b{r},\b{r})$ is the correlation part of the on-top pair density associated with the Coulomb interaction.

We can now derive the low-density limit of $\bar{E}_\c^{\sr,\mu,\l}[n]$ using the scaling relation in Eq.~(\ref{Ecsrmlscaling}) and the asymptotic expansion as $\mu\to\infty$ in Eq.~(\ref{Ecmulxmuinfty})
\begin{eqnarray}
\bar{E}_\c^{\sr,\mu,\lambda}[n_\gamma] &=& \gamma^2 \bar{E}_\c^{\sr,\mu/\gamma,\lambda/\gamma,1/\gamma}[n]
\nonumber\\
&\isEquivTo{\gamma \to 0}& \gamma^3 (1-\l) \frac{\pi}{2\mu^2}  \int n_{2,\c}^{1/\gamma}[n](\b{r},\b{r}) \d\b{r} 
\nonumber\\
&\isEquivTo{\gamma \to 0}& \gamma^3 (1-\l) \frac{\pi}{4\mu^2}  \int \left[ -n(\b{r})^2 + m(\b{r})^2\right] \d\b{r}, 
\end{eqnarray}
where we have used the strong-interaction limit of the on-top pair density, $\lim_{\gamma\to 0}n_{2,\c}^{1/\gamma}[n](\b{r},\b{r}) = -n(\b{r})^2/2 + m(\b{r})^2/2 = -2 n_\uparrow(\b{r}) n_\downarrow(\b{r})$~\cite{BurPerErn-JCP-98} where $m(\b{r})$ is the spin magnetization and $n_\sigma(\b{r})$ are the spin densities ($\sigma=\uparrow,\downarrow$).
\end{widetext}



\begin{thebibliography}{10}
\bibitem{KalTou-JJJ-XX-note2}
The term ``range-separated double hybrid (RSDH)'' has already been used in
  Ref.~\onlinecite{SanCivUsvTouShaMas-JCP-15} to designate the RSH+MP2 method
  and in Refs.~\onlinecite{CorStoJenFro-PRA-13,CorFro-IJQC-14} to generically
  designate several kinds of range-separated MP2/DFT hybrids. In the present
  work, we use RSDH to designate the specific scheme corresponding to
  Eq.~(\ref{EmulRSDH}).
\end{thebibliography}
\end{document}